\documentclass{IEEEoj}
\usepackage{cite}
\usepackage{amsmath,amssymb,amsfonts}
\usepackage{algorithmic}
\usepackage{graphicx,color}
\usepackage{threeparttable}
\usepackage{booktabs}
\usepackage{amsmath}
\usepackage{multirow}
\usepackage{subcaption}
\usepackage{textcomp}
\usepackage[hidelinks]{hyperref}
\usepackage{glossaries}

\usepackage{comment}
\usepackage{tabularx}

\makeatletter
\def\subsubsection{\@startsection{subsubsection}{3}{\z@}{-13pt \@plus -2pt \@minus -2pt}{1pt}{\subsubsectionfont}}
\makeatother

\newacronym{QML}{QML}{Quantum Machine Learning}
\newacronym{QI}{QI}{Quantum Inspired}
\newacronym{TN}{TN}{Tensor Networks}
\newacronym{QC}{QC}{Quantum Computing}
\newacronym{NISQ}{NISQ}{Noisy Intermediate-Scale Quantum}
\newacronym{QUBO}{QUBO}{Quadratic Unconstrained Binary Optimization}
\newacronym{STN}{STN}{Singularity\texttrademark\ Tensor Network}
\newacronym{SML}{SML}{Singularity\texttrademark\ Machine Learning}
\newacronym{MASE}{MASE}{Mean Absolute Scaled Error}
\newacronym{MAPE}{MAPE}{Mean Absolute Percentage Error}
\newacronym{SQER}{SQER}{Singularity\texttrademark\ Quantum-enhanced Ensemble Regressor}
\newacronym{ML}{ML}{Machine Learning}
\newacronym{MI}{MI}{Mutual Information}
\newacronym{DL}{DL}{Deep Learning}
\newacronym{TT}{TT}{Tensor Train}
\newacronym{RF}{RF}{Random Forest}
\newacronym{KNR}{KNR}{K-Neighbors Regressor}
\newacronym{SVR}{SVR}{Support Vector Regression}
\newacronym{ANN}{ANN}{Artificial Neural Network}
\newacronym{QoE}{QoE}{Quality of Experience}
\newacronym{KQI}{KQI}{Key Quality Indicator}
\newacronym{KPI}{KPI}{Key Performance Indicator}
\newacronym{maxvol}{MaxVol}{maximum volume}
\newacronym{PRB}{PRB}{Physical Resource Block}

\newacronym{AI}{AI}{Artificial Intelligence}

\newacronym{CG}{CG}{Cloud Gaming}
\newacronym{CRAN}{CRAN}{Cloud RAN}

\newacronym{eBPF}{eBPF}{extended Berkeley Packet Filter}

\newacronym{FPS}{FPS}{Frames Per Second}

\newacronym{GPU}{GPU}{Graphics Processing Units}

\newacronym{IoT}{IoT}{Internet of Things}

\newacronym{LSTM}{LSTM}{Long Short-Term Memory}
\newacronym{LTE}{LTE}{Long-Term Evolution}

\newacronym{NearRTRIC}{Near-RT RIC}{Near-real-time RIC}
\newacronym{NonRTRIC}{Non-RT RIC}{Non-real-time RIC}

\newacronym{OCU}{OCU}{ORAN Centralized Unit}
\newacronym{ODU}{ODU}{ORAN Distributed Unit}
\newacronym{ORU}{ORU}{ORAN Radio Unit}
\newacronym{ORAN}{ORAN}{Open Radio Access Network}

\newacronym{QKD}{QKD}{Quantum Key Distribution}
\newacronym{QrApp}{Q-rApp}{Quantum rApp}
\newacronym{QxApp}{Q-xApp}{Quantum xApp}

\newacronym{RAN}{RAN}{Radio Access Network}
\newacronym{rApp}{rApp}{RAN Application}
\newacronym{RIC}{RIC}{RAN Intelligent Controller}
\newacronym{RTT}{RTT}{Round-Trip Time}

\newacronym{SINR}{SINR}{Signal-to-Interference-plus-Noise Ratio}
\newacronym{SMO}{SMO}{Service Management and Orchestration}

\newacronym{UE}{UE}{User Equipment}

\newacronym{V2X}{V2X}{Vehicle-to-everything}

\newacronym{xApp}{xApp}{eXtended Application}

\glsdisablehyper

\AtBeginDocument{\definecolor{ojcolor}{cmyk}{0.93,0.59,0.15,0.02}}
\def\OJlogo{\vspace{-4pt}\hskip-4pt\includegraphics[height=18pt]{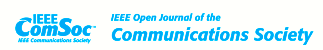}}

\begin{document}

\receiveddate{This work has been submitted to the IEEE for possible publication. Copyright may be transferred without notice, after which this version may no longer be accessible}



\title{Quantum-based QoE Optimization in Advanced Cellular Networks: Integration and Cloud Gaming Use Case}

\author{Fatma Chaouech\textsuperscript{\textsection}$\mathbf{^{,2}}$}
\author{Javier Villegas\textsuperscript{\textsection}$\mathbf{^{,1}}$}
\author{Ant\'onio Pereira\textsuperscript{\textsection}$\mathbf{^{,2}}$}
\author{Carlos Baena$\mathbf{^1}$}
\author{Sergio Fortes$\mathbf{^1}$}
\author{Raquel Barco$\mathbf{^1}$}
\author{Dominic Gribben$\mathbf{^2}$}
\author{Mohammad Dib$\mathbf{^4}$}
\author{Alba Villarino$\mathbf{^2}$}
\author{Aser Cortines$\mathbf{^3}$}
\author{Rom\'an Or\'us$\mathbf{^2}$}
\affil{Telecommunication Research Institute (TELMA), Universidad de Málaga, E.T.S. Ingeniería de Telecomunicación, Bulevar Louis Pasteur 35, 29010, Málaga (Spain)}

\affil{Multiverse Computing, Paseo de Miram\'on 170, 20014 San Sebasti\'an, Spain}

\affil{Multiverse Computing, rue de la Croix Martre, 91120 Palaiseau, France}

\affil{Multiverse Computing, 192 Spadina Ave, 509 Toronto, Canada}


\corresp{CORRESPONDING AUTHOR: Sergio Fortes (e-mail: sfr@ic.uma.es).}
\authornote{This work has been partially funded by: Ministerio de Asuntos Económicos y Transformación Digital and European Union - NextGenerationEU within the framework ``Recuperación, Transformación y Resiliencia y el Mecanismo de Recuperación y Resiliencia'' under the project MAORI. 
This work has been also supported by  Ministerio de Ciencia y Tecnología through grant FPU21/04472.}

\markboth{Preparation of Papers for IEEE OPEN JOURNALS}{Author \textit{et al.}}

\begin{abstract}
This work explores the integration of Quantum Machine Learning (QML) and Quantum-Inspired (QI) techniques for optimizing end-to-end (E2E) network services in telecommunication systems, particularly focusing on 5G networks and beyond. The application of QML and QI algorithms is investigated, comparing their performance with classical Machine Learning (ML) approaches. The present study employs a hybrid framework combining quantum and classical computing leveraging the strengths of QML and QI, without the penalty of quantum hardware availability. This is particularized for the optimization of the Quality of Experience (QoE) over cellular networks. The framework comprises an estimator for obtaining the expected QoE based on user metrics, service settings, and cell configuration, and an optimizer that uses the estimation to choose the best cell and service configuration. Although the approach is applicable to any QoE-based network management, its implementation is particularized for the optimization of network configurations for Cloud Gaming services. Then, it is evaluated via performance metrics such as accuracy and model loading and inference times for the estimator, and time to solution and solution score for the optimizer. The results indicate that QML models achieve similar or superior accuracy to classical ML models for estimation, while decreasing inference and loading times. Furthermore, potential for better performance is observed for higher-dimensional data, highlighting promising results for higher complexity problems. Thus, the results demonstrate the promising potential of QML in advancing network optimization, although challenges related to data availability and integration complexities between quantum and classical ML are identified as future research lines.
\end{abstract}

\begin{IEEEkeywords}
Quantum, Quantum Machine Learning (QML), Quantum Inspired Algorithms (QIA), Quality of Experience, Network management, 5G, 6G, Open RAN, Network intelligence, AI, optimization \vspace{-12pt}
\end{IEEEkeywords}


\maketitle

\begingroup\renewcommand\thefootnote{\textsection}
\footnotetext{These authors contributed equally to this work.}
\endgroup

\section{Introduction}
\label{section:introduction}


\IEEEPARstart{T}{he} transition from \gls{LTE} to 5G represented a major leap in cellular network technology, driven by the need for faster speeds, lower latency, and enhanced connectivity.
These needs are derived from the increase in the demand for services and new applications, such as streaming services, \gls{IoT} devices, and real-time applications such as augmented reality. This makes it ideal for mission-critical communications, autonomous vehicles, and smart cities. Key innovations include the use of millimeter wave bands for higher bandwidth, Massive MIMO for enhanced coverage, and network slicing for tailored services. Despite its benefits, the transition has faced challenges such as significant infrastructure investments, seamless handover between \gls{LTE} and 5G networks, and uneven global deployment. Furthermore, security remained a priority, with advanced protocols ensuring data protection in this high-speed environment.

The variety of services and requirements has increased the complexity of networks that, together with the need to handle large amounts of data from services and devices, have made \gls{ML} a crucial tool for managing modern cellular networks. In this way, by using algorithms \gls{ML}, networks can automate maintenance tasks, such as coverage optimization or interference reduction, by dynamically adjusting the configurations of cell towers without human intervention. Additionally, \gls{ML} can enhance traffic management by predicting congestion peaks and efficiently rerouting or prioritizing data, ensuring smoother performance during high demand. It also allows personalized user experiences by analyzing behavior and tailoring services. Moreover, \gls{ML} might strengthen network security by identifying unusual patterns that can indicate threats, allowing proactive measures to protect against attacks.

Furthermore, \gls{ML} helps to futureproof networks by anticipating technological advancements and adapting the infrastructure accordingly. All these \gls{ML} applications reduce operational costs by automating routine tasks such as maintenance scheduling and resource optimization, allowing providers to focus on improving the infrastructure and developing new services. Hence, \gls{ML} is essential for managing the intricacies of cellular networks, optimizing performance, enhancing security, personalizing services, cutting costs, and ensuring adaptability to future developments.

Classical \gls{ML} is currently applied in many areas of cellular network management, such as corrective and predictive maintenance, network security, resource allocation, and personalization of user experience. Corrective and predictive maintenance uses \gls{ML} to detect and identify and anticipate failures, respectively, to take corrective measurements. For instance, the authors in \cite{Sundsvist2023Bottleneck} developed a method to detect bottlenecks in 5G \gls{RAN} to avoid failures. Furthermore, the authors in \cite{Kawasaki2023eBPFPrediction} proposed a \gls{LSTM}-based prediction system powered by \gls{eBPF} to predict failures in 5G cloud core networks.

Regarding network security, \gls{ML} is applied to detect unusual traffic patterns that can be indicative of threats or cyberattacks. In this area, \gls{ML} is crucial to provide increased security for safety-critical applications of cellular networks, such as \gls{V2X} and industrial communications. In this area, the authors of \cite{Dey2025HOV2X} proposed a model based on \gls{ML} to detect and localize rogue vehicles trying to impede handovers in \gls{V2X} scenarios.

In the area of resource allocation, the \gls{ML} models can be used to optimize how resources such as bandwidth and power are distributed among the network to maximize performance and efficiency. For example, the authors in \cite{Zhao2024ResAllocRL} developed transformer-based deep reinforcement learning for multislot multiuser resource allocation, resulting in increased spectral efficiency and scheduling fairness. Moreover, \gls{ML} can also be applied towards user experience personalization, tailoring network configuration to enhance user satisfaction individually based on the quality indicators of the services being consumed. Here, the authors in \cite{Zhang2024ResAllocQoE} proposed a technique for allocating network resources with the aim of achieving a trade-off between the capacity of the system and \gls{QoE}.


\gls{QML} offers a promising alternative by applying the quantum mechanics principle, such as superposition and entanglement, to explore high-dimensional data spaces more efficiently. Although current quantum hardware remains in early development, \gls{QI} \gls{TN} \cite{Orús2019} have been shown to efficiently model correlations in high-dimensional systems, making them potentially useful for cellular network management tasks. Furthermore, hybrid quantum-classical approaches further bridge the gap between classical and quantum computing, allowing the integration of such techniques into real-world applications. Furthermore, as cellular networks advance further, the ability to process and optimize large amounts of data in real time will be critical.

With the objective of assessing the applicability of quantum approaches to advanced cellular networks, this article explores the potential of \gls{QML} and \gls{QI} methods in cellular network management, covering their applications in metrics estimation while highlighting key challenges and future research directions. In this way, the main contributions of this works is the development and description of a framework that enables the integration of Quantum and \gls{QI} \gls{ML} for estimation of metrics and the optimization of services. Furthermore, the present work tackles a specific use case for the optimization of \gls{QoE} in a \gls{CG} service, comparing the results of the quantum-based approaches against classical \gls{ML} methods.

For doing so, this article follows the following structure. Section~\ref{sec:QML} describes the principles, state-of-the-art, and potential benefits in mobile networks of \gls{QML} and \gls{QI}. Then, Section~\ref{sec:ORAN} proposes a system for optimizing network configurations based on services \gls{QoE} with \gls{QML} and \gls{QI} algorithms and introduces a possible scheme for integrating it into cellular networks taking advantage of the \gls{ORAN} architecture. Section~\ref{sec:usecase} proposes a case study for applying \gls{QML} to manage \gls{QoE} of a \gls{CG} service offered on top of the cellular network and Section~\ref{sec:eval} presents the results of the proposed approach. Lastly, Section~\ref{sec:conclusions} summarizes the key results, limitations, challenges, and opportunities of \gls{QML} in cellular networks.  

\section{Quantum and Quantum-Inspired Machine Learning}\label{sec:QML}

\gls{QC} is an alternative computational model based on the principles of quantum mechanics. Unlike classical bits, which encode information as either 0 or 1, qubits can exist in superposition, meaning they simultaneously represent multiple states. Additionally, entanglement enables strong correlations between qubits, allowing quantum algorithms to manipulate information in ways that do not have a classical equivalent \cite{Nielsen_Chuang_2010}.

Although large-scale, fault-tolerant quantum computers remain a long-term goal, current \gls{NISQ} devices already offer potential advantages in optimization and simulation problems \cite{Preskill_2018}. Among quantum computing models, gate-based quantum computing is universal, meaning it is capable of running any quantum algorithm that is theoretically computable, while adiabatic quantum computing \cite{RevModPhys.90.015002} is particularly suited for combinatorial optimization, often formulated as \gls{QUBO}\cite{10.3389/fphy.2014.00005} problems.

Some of the techniques used in quantum computing, particularly those used for optimization and simulation, have inspired their application to \gls{ML}, where many tasks can leverage large-scale data processing and high-dimensional optimization. \gls{QML} investigates how quantum computing can improve data-driven learning tasks by providing alternative methods to represent and transform complex information \cite{Biamonte2017}. However, integrating quantum computation into machine learning requires careful consideration of how classical data is encoded into quantum states and how meaningful results can be extracted.

Given that large-scale, fault-tolerant quantum processors are still under development, hybrid quantum-classical models currently offer the most practical approach. In these models, quantum circuits handle specialized subroutines — such as encoding non-trivial feature spaces or optimizing certain objective functions — and their outputs are fed into classical algorithms that handle parameter optimization and model updating \cite{PhysRevA.101.032308}.

As research progresses, \gls{QML} continues to evolve alongside advancements in quantum hardware, data encoding techniques, and hybrid computational strategies. While practical advantages over classical methods are still being explored, quantum-enhanced learning models offer a promising framework for integrating quantum and classical computational techniques.

In parallel to these developments, and separate from \gls{QML}, \gls{QI} algorithms adapt quantum principles for use on classical computers, delivering practical benefits without the need for quantum hardware. Techniques like \glspl{TN} \cite{Orús2019}, originally developed in quantum physics to efficiently model many-body systems, are now leveraged in data science to manage high-dimensional datasets. \glspl{TN} excel in modeling complex relationships while significantly reducing computational resource requirements, making them a powerful tool for optimization and predictive modeling in scenarios where conventional methods may falter \cite{Sengupta2022-yd}.

Classical \gls{ML} and \gls{DL} typically represent input data as real-valued vectors in fixed-dimensional Euclidean spaces. Features extracted from structured data, images, or text are directly processed without transformation into alternative computational representations. In contrast, \gls{QML} encodes classical data into quantum states using quantum feature maps, embedding it into high-dimensional Hilbert spaces. This quantum encoding can, in principle, provide access to non-trivial geometries or correlations that may be inaccessible to classical methods.

Optimization strategies also differ substantially. \gls{DL} models are trained using gradient-based optimization, most commonly through backpropagation applied to networks of differentiable functions. Classical \gls{ML} algorithms often avoid gradients altogether, relying instead on techniques such as greedy heuristics, convex solvers, or closed-form solutions. In gate-based \gls{QML}, optimization typically involves a hybrid quantum-classical loop, where a classical optimizer updates the parameters of Parameterized Quantum Circuits (PQCs) based on measurement outcomes from quantum hardware or simulators. Gradients, when required, are estimated using quantum-aware techniques such as the parameter-shift rule or finite-difference methods. In Adiabatic Quantum Computing (AQC), by contrast, optimization is performed via the adiabatic evolution of a quantum system toward the ground state of a Hamiltonian encoding the objective function. This gradient-free process has been used for tasks such as feature selection, binary classification, and clustering.

Regarding their role in telecommunication, quantum technologies can be applied in a variety of areas, such as data transmission, routing, and security \cite{Hildebrad2023QuantumComms}. Here, the most prominent application of quantum for communications is in security, where \gls{QKD} \cite{BENNETT20147QKD} is discussed to ensure message privacy against Post-Quantum Cryptography attacks. However, quantum computing can also be applied to other tasks, such as baseband signal processing for base stations \cite{Srikar2023QuantumBBU}, which could replace current technology but offer much lower power consumption and therefore reduced operation costs. Lastly, \gls{QML} is already being studied to solve some classical optimization problems such as radio resource optimization \cite{LeTung2025QGNN}. Although there are approaches using \gls{QML} for wireless network management, their real-life application and integration are hardly explored.

\section{Proposed System and Integration in the O-RAN architecture}\label{sec:ORAN}

\begin{figure*}[t]
    \centering
    \includegraphics[width=\textwidth,clip=True, trim={0pt 880pt 0pt 870pt}]{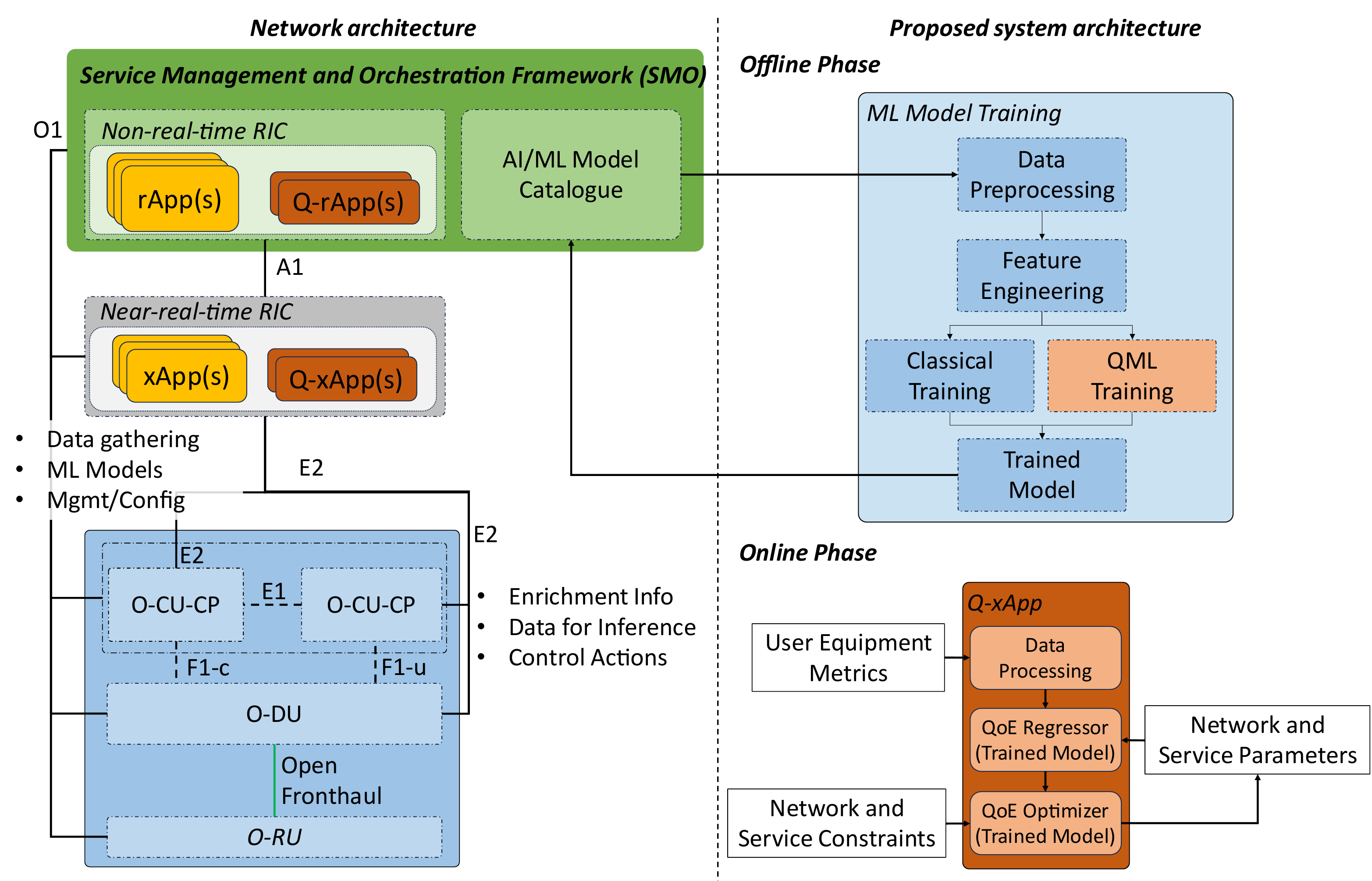}
    \caption{Network architecture and integration of QML/QI ML algorithms for a QoE Estimation and Optimization.}
    \label{fig:overview}
\end{figure*}

\gls{ORAN} is an architecture based 3GPP standard, but containing multiple extensions and core principles of interoperability, virtualization, and modularity that allow for a flexible and cost-effective approach to networking by decoupling functions from specific hardware and utilizing open standards. Moreover, this architecture is an evolution from \gls{CRAN}, which aimed to reduced network operation cost and increased flexibility by separating base station functionalities and offloading signal processing to the cloud, centralizing the computation load and allowing better resource sharing. In \gls{ORAN}, base stations are segmented into 3 components, the \gls{OCU}, \gls{ODU} and \gls{ORU}. The \gls{OCU} can be deployed further in the cloud as it is responsible of performing the upper layers tasks and, therefore, it has lesser latency requirements. Conversely, the \gls{ODU} needs to be deployed closer to the \gls{ORU}, as it is responsible for the lower layers tasks, which demands lower latencies.

Furthermore, \gls{ORAN} contemplates \gls{ML} and \gls{AI} as part of the management of the \gls{RAN}. For this, it defines the \gls{RIC}, a component where algorithms and applications can connect to perform management tasks on the \gls{RAN}. There are two flavours of the \gls{RIC}, the \gls{NearRTRIC} and the \gls{NonRTRIC}. The \gls{NearRTRIC} is directly connected to the distributed elements of the \gls{RAN} and algorithms are deployed in the form \glspl{xApp}, which control near-real-time events (i.e., from 10 milliseconds to 1 second). Regarding the \gls{NonRTRIC}, it is deployed within the \gls{SMO} and communicates with the \gls{NearRTRIC} through \glspl{rApp} to provide policy-based guidance, managing non-real-time events (i.e., more than 1 second).

Although \glspl{rApp} and \glspl{xApp} are logically part of the \gls{NonRTRIC} and \gls{NearRTRIC}, respectively, all the components are also decoupled. This means that the algorithms can be run on different host machines than the \glspl{RIC}, allowing them to be run on machines with specific hardware to cover their processing needs (e.g., \glspl{GPU} for \gls{AI}/\gls{ML}). This flexibility eases the integration of new technologies, such as \gls{QC}, into the \gls{ORAN} architecture, which presents an interesting opportunity to enhance the management of the cellular network. In Figure~\ref{fig:overview}, \gls{QML} and \gls{QI} algorithms are integrated into this framework in the form of \glspl{QrApp} and \glspl{QxApp} in the \gls{NearRTRIC} and \gls{NonRTRIC}, respectively, following the architecture proposed in \cite{QuantumORANMagazine}. In this instance, these applications refers to both, \gls{QML} and \gls{QI} algorithms signifying the quantum nature of the algorithms, regardless of being run on quantum hardware or not.

Extending this framework to incorporate \gls{QML} and \gls{QI} algorithms is a logical evolution that could leverage \gls{QC}'s potential for performance improvements. In this regards, the authors in \cite{EricssonQuantum} made a proposal for the deployment of quantum computers interconnected with classical computers in the cloud. However, some \gls{QML} methods require quantum hardware for execution or training, raising practical concerns regarding the feasibility of implementation, especially given the specialized and costly nature of such hardware. Conversely, \gls{QI} algorithms can operate entirely on classical hardware, offering a more immediate and practical solution. Additionally, as another stepping stone towards fully quantum-based \gls{AI}/\gls{ML} in cellular network management, some \gls{QML} methods only need quantum hardware for training and can run on classical systems afterward. Furthermore, although \gls{QC} hardware is still expensive to use at scale given that it is not yet mature enough, \gls{QML} algorithms for \glspl{QxApp} and \glspl{QrApp} proof-of-concepts by running the applications in the network for gathering metrics and applying decisions, but offloading the quantum processing to a quantum cloud following a request/response scheme. This could potentially introduce issues like higher latency and security risks due to data leaving the network, limiting its application in real live network and for algorithms whose latency requirements are above tens of milliseconds.

The prospect of enhanced processing efficiency through quantum parallelism is deemed crucial, as it could significantly benefit tasks such as classification, clustering, regression, and optimization in network management. Therefore, addressing these challenges requires a strategic approach, possibly involving a hybrid model in which \gls{QI} algorithms are used in the near term due to their compatibility with the existing infrastructure, while \gls{QML} methods are gradually incorporated as quantum hardware becomes more accessible. Moreover, ensuring compatibility within the \gls{ORAN} architecture without compromising its open standards is crucial for widespread adoption.

Furthermore, exploring how these new applications interact with the existing framework and whether they require novel interfaces or modifications is essential. The potential for optimization in resource allocation and traffic management is promising, but concrete examples or case studies that demonstrate successful real-world applications should provide valuable insights into their impact. 

Consequently, Figure~\ref{fig:overview} also shows the proposed system for building and deploying quantum-based mechanisms for the optimization of services running on top of the cellular network. Here, two different phases can be distinguished, the offline phase and the online phase, for the model training phase and the deployment for operation, respectively. 

Firstly, the offline phase corresponds to a general training of a model that could require a classical or quantum traning process. For this, data stored at the \gls{SMO} is pre-processed and feature engineering is performed to select the most relevant features. The model is then trained on a classical computer or on a quantum computer, depending on the requirements of the training. Lastly, the trained model is stored in the \gls{SMO} as part of the catalog of \gls{AI}/\gls{ML} models for posterior use.

Secondly, the online phase corresponds to the real operation of the models which, in this instance, are deployed in the form of a \gls{QxApp}. This \gls{QxApp} first gathers metrics from \gls{UE} and transforms them for proper use in the model. Then, the \gls{QoE} regressor obtains the quality of the current \gls{UE} signal quality and the configuration of network and services. Lastly, this \gls{QoE} is used as input for the optimizer along with the network and service configuration constraints to the new network configuration proposed. In this way, the proposed system can take into account the metrics \gls{UE} and the constraints of the network and the service to provide the network configuration that maximizes \gls{QoE}.

In this way, this work proposes a system for optimizing network configurations based on the \gls{QoE} for the services that users are consuming. Furthermore, the integration of said system is also described considering the \gls{ORAN} network architecture. Here, the integration of \gls{QML} and \gls{QI} algorithms into \gls{ORAN} offers exciting possibilities to advance network management capabilities, but careful consideration of hardware limitations, latency, security, and practical deployment strategies is essential to fully realize its potential.

\section{Case Study: QML-Powered QoE Estimation and Optimization for Cloud Gaming Services}\label{sec:usecase}
As a case study for the use of the proposed \gls{QML} system in the management of the cellular network, shown in Figure~\ref{fig:overview}, the optimization of a \gls{CG} service is proposed. A \gls{CG} service is a platform that allows users to play high-quality video games on various devices, such as smartphones, tablets, PCs, or smart TVs, without the need for expensive or specialized hardware. For this,instead of running video games directly on the user's device, the processing is done remotely on powerful servers located in data centers. Then, the game is streamed over the internet to the user's device, where it can be played in real time, thus, eliminating the need for users to purchase consoles of high-end computers. Furthermore, \gls{CG} allows seamless gameplay across multiple devices, allowing players to start a game on one device and continue it on another without interruption.

In this way, the key benefits of \gls{CG} services include instant access to new releases, reduced hardware-related issues, and enhanced mobility for gamers. However, \gls{QoE} depends heavily on the speed and stability of the internet, as high latency or poor connections can lead to lag, buffering, and/or reduced video resolution. Despite these challenges, \gls{CG} represents a significant shift in how games are delivered and consumed, offering a flexible and convenient way to enjoy high-quality gaming experiences. 

In order to offer a satisfactory experience to \gls{CG} users, it is important that a good level of \gls{QoE} is met. However, classical network metrics such as counters and \glspl{KPI} do not offer a clear view of the \gls{QoE}, instead, metrics derived from the service are required. The metrics from the service are denominated \glspl{KQI} \cite{KQIModeling} and provide a measurement of the service performance that can directly affect the perceived \gls{QoE}.

In this case, the \gls{QoE} is mainly influenced by \gls{CG} metrics, such as resolution, \gls{FPS} and latency. Thus, a framework which estimates a \gls{CG} session's \gls{QoE} from network metrics and service configuration is proposed. Then, this estimator can be used to feed an optimizer that will search for a near-optimal configuration, that satisfies the minimum quality requirement, and in addition will aim to reach the best trade-off between perceived quality and network resources usage, using a parametrized cost function for the optimizer.

\subsection{
QoE Estimation} 



\subsubsection{Dataset Description}
In order to train and evaluate the \gls{QML} and \gls{QI} regression models, the end-to-end dataset from \cite{k0w8-qz67-22}, which includes the \glspl{KQI} that are desired to estimate, is leveraged. This dataset contains a total of $8,461$ samples, spanning \textit{Video on Demand} ($2,784$), \textit{Live Streaming} ($1,933$), and \textit{Cloud Gaming} ($3,694$); collected over \gls{LTE} ($5,413$), 5G ($1,344$), Ethernet and WiFi ($1704$). Its 252 features (43 categorical, 209 numerical) capture both configurable parameters (resolution, frame rate) set at the server level and network \glspl{KPI} (\gls{SINR}, \gls{RTT}, packet loss), providing the rich, multi-modal input required to assess the models. These samples were obtained in a 4G/5G controlled testbed. 

The data collection, as described in \cite{10148946}, was conducted in a controlled testbed environment that simulates real-world conditions within 4G and 5G networks. This setup enabled the precise measurement of \glspl{KQI}, which serve as proxies for the \gls{QoE} perceived by end users. To ensure dataset consistency regarding network impact, all samples were gathered under a standardized service configuration and execution framework, ensuring uniformity in service delivery and evaluation.





\subsubsection{Data Preprocessing}\label{sec:data_processing}

Since this work focuses on the estimation and optimization of \gls{CG}, only the $3694$ entries relating to this service were selected. Moreover, to ensure data quality and streamline model training, a structured preprocessing pipeline was applied. First, outliers were detected and removed through statistical analysis. Then , the categorical variable \textit{Resolution}, initially expressed as discrete labels (e.g., 720p, 1080p, 4K), was encoded into numerical values (0, 1, 2, 3). Next, the dataset was split into training (70\%), validation (15\%), and test (15\%) sets to ensure unbiased model evaluation. 


The next step for the preparation of the data is applying feature normalization. However, the normalization process depends on the chosen model and, therefore, further details will be provided when defining the different approaches.

After applying the detailed preprocessing procedure, the dataset was refined to 3,467 samples and 16 features, including both predictor input variables and target \glspl{KQI}.

\subsubsection{Feature Selection}
To determine the most informative features for predicting each \gls{KQI}, the \gls{MI}\cite{InformationTheory} score was calculated. \gls{MI} quantifies the dependency between the features and the target variables, capturing both linear and nonlinear relationships. A \textbf{higher} \gls{MI} score implies that the feature is \textbf{more informative} for predicting the target variable, as it captures significant dependencies between the two variables. The \gls{MI} score is calculated as
\begin{equation}
I(X, Y) = \sum_{i=1}^{n} \sum_{j=1}^{m} P(x_i, y_j) \log \left( \frac{P(x_i, y_j)}{P(x_i) P(y_j)} \right),
\end{equation}
where $I(X,Y)$ is the \gls{MI} between metrics $X$ and $Y$, $P(x_i, y_j)$ is the joint probability mass function between the metrics, and $P(x_i)$ and $P(y_j)$ are the marginal probability mass functions of metrics $X$ and $Y$, respectively.

\begin{figure}[t]
    \centering
    \includegraphics[width=\columnwidth]{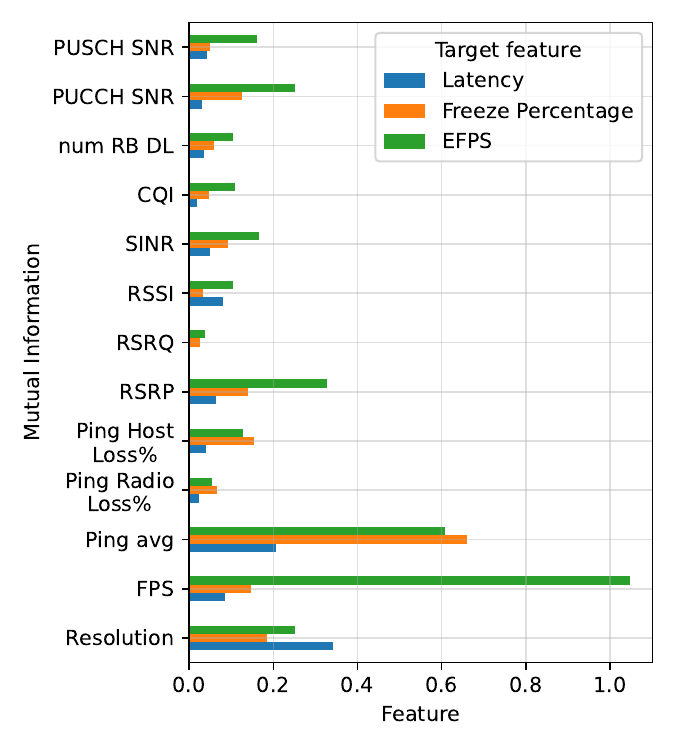}
    \caption{Mutual information scores for each \gls{CG} performance metric. Key factors include FPS, Ping avg, and Resolution.}
    \label{fig:mi_scores_triangle}
\end{figure}

Accordingly, the \gls{MI} scores have been obtained for all input features and the \gls{CG} target \glspl{KQI} of the proposed dataset, presented in Figure~\ref{fig:mi_scores_triangle}. Firstly, the figure shows that the most relevant features for latency are frame rate (\textit{FPS}), average ping (\textit{Ping avg}), and resolution.

Secondly, for the freeze percentage \gls{KQI}, \textit{Ping avg} stands out with the highest score. It is followed by \textit{Ping Host Loss} and \textit{FPS}, suggesting that freeze events are primarily affected by network conditions.

In the case of effective frames per second (\textit{EFPS}), fps again provides the highest \gls{MI} score. \textit{Ping avg} and \textit{RSRP} follow, indicating that both network delay and signal strength contribute moderately to the output. Traditional radio metrics such as \textit{SINR}, \textit{RSRQ}, and \textit{RSSI} show low scores across all KPIs, underscoring their limited relevance in this context.

\subsubsection{Evaluation Function}
Before selecting a training approach, it is essential to define a method for quantifying error. The evaluation function should be model-intrinsic, as it serves to compare different approaches objectively. For this reason, the \gls{MASE} was chosen as the evaluation metric.

Unlike traditional error metrics, \gls{MASE} measures model performance relative to a simple baseline, providing a clear indication of whether the model improves upon a naive estimation. It also offers key advantages over the \gls{MAPE}, as it applies the same penalty to both positive and negative errors, ensuring a more balanced assessment. Additionally, \gls{MASE} avoids the issue of infinite or undefined values, which can arise in \gls{MAPE} and symmetric \gls{MAPE} (sMAPE), when dealing with zero or near-zero values in the dataset. Given that some \glspl{KQI} in \gls{CG}, such as Freeze Percentage, can take values close to zero, \gls{MASE} is particularly well-suited for evaluating model accuracy in this context.

\subsection{Quantum and Quantum-inspired ML implementation for KQI estimation}

For the estimation of the KQIs, both a QMl and a QI approach are considered as detailed in the next sections. 

\subsubsection{QML approach}

Current quantum hardware faces several challenges that limit its practical integration in communication networks. These constraints include a limited number of qubits, high error rates, and the need for expensive and specialized cryogenic setups, all of which result in high operational costs and complex maintenance requirements. Such limitations not only hinder scalability but also impede the reliable execution of quantum algorithms in dynamic, high-demand environments.

To address these challenges, the \gls{QML} approach adopted in this work is based on adiabatic quantum computing. In particular, the \gls{SQER}, part of Multiverse Computing’s \gls{SML} framework, was used. \gls{SQER} is derived from the QBoost algorithm \cite{pmlr-v25-neven12}, which performs model training via quantum annealing. A key advantage of this approach is that inference does not require quantum hardware, allowing the trained model to be deployed using standard classical infrastructure.

Originally designed for binary classification \cite{neven2009traininglargescaleclassifier}, QBoost constructs an ensemble of weak learners by formulating their selection as a \gls{QUBO} problem \cite{glover2019tutorialformulatingusingqubo}. This formulation enables the use of quantum annealers to select the subset of learners that optimizes classification performance.

QBoost has since been extended to regression tasks \cite{góes2021qboostregressionproblemssolving}. In this setting, each weak learner is assigned a floating-point weight, encoded as a set of binary variables using a floating-point expansion technique \cite{rogers2019floatingpointcalculationsquantumannealer}. After optimization, the binary representation is decoded to obtain the final weights, enabling continuous-valued predictions.

This method does not require specialized feature normalization techniques. Instead, a standard min-max scaling is applied:

\begin{equation} x_{scaled} = \frac{x_i - x_{min}}{x_{max} - x_{min}} \end{equation}

This transformation ensures that all input variables fall within a comparable range, mitigating scale-related biases during training.

\subsubsection{Quantum-Inspired approach}

In parallel, the \gls{STN} Regressor is a quantum-inspired method that leverages tensor networks and the principles of Exponential Machines \cite{novikov2017exponentialmachines} to address regression problems by modeling the complete set of $2^N$ interactions among $N$ features. By optimizing the weights of these interactions within the tensor network structure, the \gls{STN} regressor learns effectively from high-dimensional feature spaces while mitigating the computational challenges that arise from modeling an exponential number of feature interactions. 

Due to the discrete nature of tensor network representations, they require input features to be represented as integer indices, meaning input features must be normalized differently from standard \gls{ML} tasks. For this, first each feature's minimum and maximum allowed values, $x_{max\_allowed}$ and $x_{min\_allowed}$, respectively, are defined  based on predetermined standardized network operation ranges. For discrete features, the number of allowed values is given by $n_{allowed\_vals} = x_{max\_allowed}$ - $x_{min\_allowed}$; for continuous features, $n_{allowed\_vals}$ was fixed at 100. Once all the scaling parameters are defined, the adapted min-max scaling is applied as
\begin{equation}
x_{scaled} = \frac{x_i - x_{min\_allowed}}{x_{max\_allowed}-x_{min\_allowed}} \times n_{allowed\_vals}.
\end{equation}

After scaling, each $x_{scaled}$ is rounded to the nearest integer, while ensuring it falls within the interval $[0,n_{allowed\_vals})$.

Once the features are normalized and discretized, the \gls{TN} structure is constructed using a \gls{TT} format. Here, \gls{MI} score obtained from feature selection guides the placement of features: those with the highest \gls{MI} scores will be kept in the center of the structure, while those with lower \gls{MI} scores are placed closer to the edges.

During training, the Adam optimizer was used, and it produced stability issues similar to those encountered with standard Stochastic Gradient Descent (SGD) when not using specialized Riemannian optimization techniques, as noted in \cite{novikov2017exponentialmachines}. It is important to note that although Riemannian optimization can improve training stability, it may also increase the risk of overfitting in scenarios where intricate, non-linear feature interactions dominate.



\subsection{Optimization use case}

Effective cellular network management relies heavily on optimization techniques, which automatically configure various network parameters to enhance performance, minimize costs, and ensure consistent user experiences.

In the context of \gls{CG}, where providers must sustain consistent frame rates (e.g. $60$ FPS) and maintain certain resolutions (e.g. $1080$p) under dynamic load, optimization is particularly important for dynamically allocating resources, balancing network load, and mitigating interference to meet these stringent quality requirements. With the purpose of providing a satisfactory \gls{QoE} while maintaining reasonable network resource spending, the optimizer objective function has been formulated taking into account the \gls{PRB} allocation and the \glspl{KQI} values. The objective function is then defined as
\begin{equation}
    J = \alpha \cdot F_S(KQIs) + (1-\alpha)\cdot F_N(PRB),
\end{equation}
where $F_S$ is the Service Cost function, which quantifies the impact of parameter settings on user-perceived \gls{QoE} by rewarding improvements in \glspl{KQI} such as higher FPS and resolution; $F_N$ is the Network Cost function, which evaluates the efficiency of network resource utilization by penalizing excessive allocation of \glspl{PRB}, thereby promoting solutions that meet service quality requirements at minimal network cost; and $\alpha \in [0, 1]$ is a trade-off parameter that balances network resource usage against \gls{QoE}, for instance $\alpha = 0.8$ prioritizes user experience, whereas $\alpha = 0.2$ favors resource savings.

Moreover, the Service Cost function is defined as
\begin{align}
        F_S (KQIs) &= 0.25 \cdot \frac{1}{1 + e^{-(Res - MinRes)}}\notag\\
                   &+ 0.25 \cdot \left( 1- e^{-\frac{\widehat{efps}}{43.25}}\right)\\
                   &+ 0.5 \left( e^{-0.5 \cdot \left(\frac{\widehat{latency}}{150} \right)} \right),\notag
\end{align}
where $Res$ is the chosen resolution and $MinRes$ its minimum acceptable value, set by the user, both using the values described in \ref{sec:data_processing}; $\widehat{efps}$ and $\widehat{latency}$ are \glspl{KQI}, obtained from the regression models by fixing the network condition features (\texttt{Ping avg}, \texttt{SINR}, \texttt{RSRP}, etc.), and changing the parameters that the system aims to optimize (\texttt{Resolution}, \texttt{fps}, and \texttt{\glspl{PRB}}).
 
Finally, the Network Cost function is modeled as:
\begin{equation}
    F_N(PRB) = \frac{1}{1+e^{-\frac{(PRB -53)}{10}}},
\end{equation}
which takes as its singular input the amount of \glspl{PRB}, $PRB \in \{1,\ldots,106 \}$, allocated to the user. It is noteworthy to highlight that all the cost functions are modeled using exponential functions with the end of 
represent the effect of diminishing returns, promoting improving on all categories, rather than maximizing one part, obtaining more balanced solutions.




\subsection{Quantum-inspired implementation for RAN parameters optimization}
While the problem at hand remains small enough for a simple exhaustive search, it is explored how \gls{STN} can enhance discrete optimization efficiency. The \gls{STN} optimization algorithm used is based on TTOpt \cite{TTOpt}, a gradient-free optimization algorithm designed for high-dimensional discrete problems, which efficiently approximates the objective function while minimizing the number of evaluations. 

When optimizing a function over a discrete domain, evaluating all possible function values becomes infeasible due to the curse of dimensionality. To address this, \gls{STN} represents the function using the \gls{TT} format, reducing computational costs by decomposing the optimization problem into lower-dimensional components. It iteratively selects key grid points for function evaluation using the \gls{maxvol} \cite{Goreinov2001} principle. 

This process refines the function approximation while systematically searching for the optimal value. The optimization progresses through alternating forward and backward sweeps, ensuring an efficient exploration of all dimensions while focusing on regions of interest. This structured search efficiently balances exploration and exploitation, making it particularly well-suited for high-dimensional discrete optimization problems.

The algorithm stops when a predefined condition is met, such as reaching the maximum number of function evaluations or detecting no further improvement. The final output is the grid point corresponding to the best function value found during the search.

\subsection{Setup description}

The work has been developed using a private network solution, specifically a network-in-a-box solution from Amarisoft \cite{amarisoft}, which provides a 5G Standalone core network and gNB. Then, the \gls{CG} service is deployed in a edge server connected to the network and integrated with the framework described in \cite{Baena2021}.

Singularity\texttrademark is a software platform developed and maintained by Multiverse Computing that provides both quantum and quantum-inspired solutions. With a strong focus on industrial integration, it offers core tools for optimization, \gls{LTE}, and \gls{AI}, along with specialized applications across sectors such as finance, energy, life sciences, and more. In this work, Singularity\texttrademark is used to demonstrate real-world applications of quantum and quantum-inspired \gls{ML} in cellular network management. Moreover, elements of the \gls{SML} framework have contributed to functionalities available in IBM’s quantum computing ecosystem through Qiskit, as detailed in the IBM Quantum documentation \cite{IBMSML}.


\begin{table*}[t]
    \caption{Performance metrics for best performing models for each \gls{KQI}.}
    \label{tab:kqi-performance}
    \begin{tabularx}{\textwidth}{c cccc cccc cccc}
    \toprule
    \multirow{4}{*}{\textbf{Models}} & \multicolumn{4}{c}{\textbf{Latency}} & \multicolumn{4}{c}{\textbf{Freeze Percentage}} & \multicolumn{4}{c}{\textbf{EFPS}} \\
    \cmidrule(lr){2-5} \cmidrule(lr){6-9} \cmidrule(lr){10-13}
    & \textbf{\shortstack{Loading \\ Time \\(ms)}} & \textbf{\shortstack{Inference\\ Time\\ (ms)}} & \textbf{\shortstack{\phantom{T} \\ MASE \\ \phantom{T}}} & \textbf{\shortstack{\phantom{T} \\ Feature \\ \phantom{T}}} &
        \textbf{\shortstack{Loading \\ Time \\(ms)}} & \textbf{\shortstack{Inference\\ Time\\ (ms)}} & \textbf{\shortstack{\phantom{T} \\ MASE \\ \phantom{T}}} & \textbf{\shortstack{\phantom{T} \\ Feature \\ \phantom{T}}}  &
        \textbf{\shortstack{Loading \\ Time \\(ms)}} & \textbf{\shortstack{Inference\\ Time\\ (ms)}} & \textbf{\shortstack{\phantom{T} \\ MASE \\ \phantom{T}}} & \textbf{\shortstack{\phantom{T} \\ Feature \\ \phantom{T}}}  \\

     \midrule
     \gls{SML} & $0.693$ & $7.463$ & $0.539$ & $9$ & $0.501$ & $6.099$ & $0.135$ & $9$ & $0.539$ & $9.374$ & $0.076$ & $13$\\
     \gls{STN} & $13.244$ & $68.560$ & $0.529$ & $8$ & $12.885$ & $11.384$ & $0.119$& $13$ & $15.080$ & $17.669$ & $0.109$ & $10$ \\
     \gls{SVR} & $7.891$ & $8.749$ & $0.538$ & $10$ & - & - & - & - & - & - & - & - \\
     \gls{RF}  & - & - & - & - & $7.958$ & $8.658$ & $0.139$ & $9$ & - & - & - & - \\
     \gls{KNR} & - & - & - & - & - & - & - & - & $0.789$ & $10.015$ & $0.084$& $5$ \\
    \bottomrule
    \end{tabularx}
\end{table*}

\begin{figure}[!htb]
    \centering
    \vspace{0pt}
    \begin{subfigure}[t]{0.85\columnwidth} 
    \raggedright
    \includegraphics[width=\columnwidth,trim={9pt 8pt 10pt 10pt},clip=True]{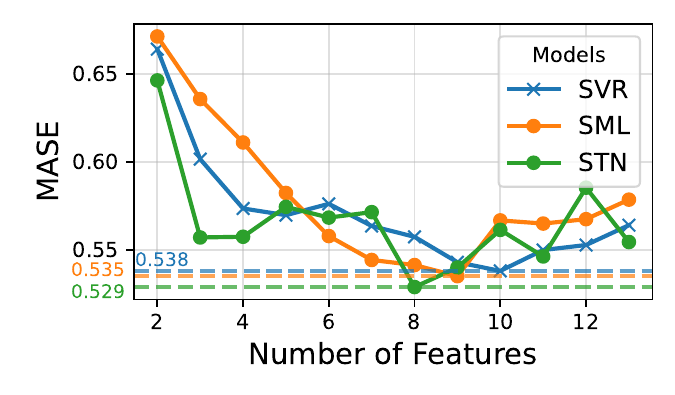} 
    \caption{Latency}
    \label{fig:best-latency}
    \end{subfigure}

    \begin{subfigure}[t]{0.85\columnwidth}
    \raggedright
    \includegraphics[width=\columnwidth,trim={4pt 9pt 10pt 10pt},clip=True]{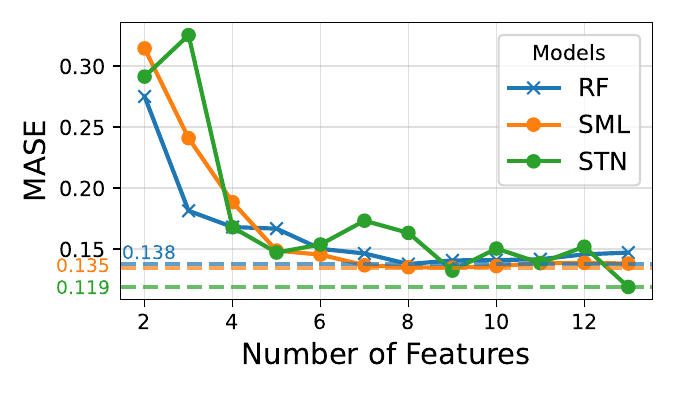} 
    \caption{Freeze Percentage}
    \label{fig:best-freeze}
    \end{subfigure}

    \begin{subfigure}[t]{0.85\columnwidth}
    \raggedright
    \includegraphics[width=\columnwidth,trim={4pt 9pt 10pt 10pt},clip=True]{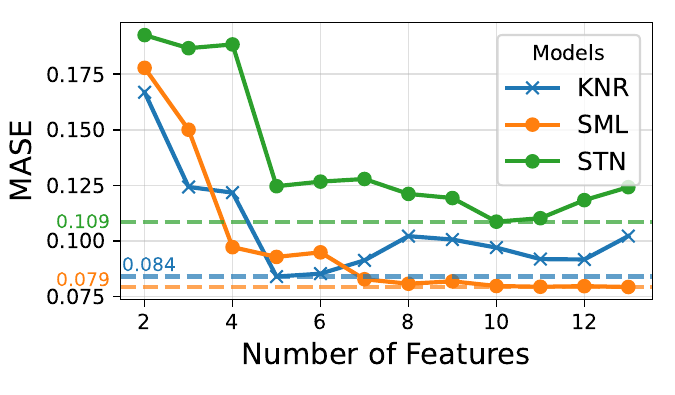}
    \caption{Effective Frames per Second}
    \label{fig:best-efps}
    \end{subfigure}

    \caption{MASE evolution of different models predicting KQIs in relation to Number of Features.}
    \vspace{-20pt}
    \label{fig:mase-best-kqi}
\end{figure}

\section{Evaluation}\label{sec:eval}

To enable a fair and consistent evaluation of the quantum-enhanced and quantum-inspired methods for predicting each \gls{KQI} against classical approaches, the benchmarking approach presented in a previous study \cite{BAENA2023109808} has been extended. From said study, the following models have been considered as baseline references:

\begin{itemize} 
\item \textbf{Latency}: \gls{ANN} were reported as the best-performing model in the original study. However, the same level of performance was unable to be reproduced in our experiments. As a result, the next-best model from the mentioned study, \gls{SVR}, was selected as the baseline for this task, being the best-performing model reproducible. 
\item \textbf{Freeze Percentage}: \gls{RF}, consistent with the top-performing model identified in the benchmark. 
\item \textbf{EFPS}: \gls{KNR}, identified as the best performer in the original benchmark. 
\end{itemize}

Moreover, in order to avoid overfitting and obtain better generalization in the results, all models were trained following five-fold cross-validation. In this way, the dataset is split into multiple portions, which are used in different iterations for training and validating with the objective of evaluating the model progress on unseen data.

\subsection{Estimation Error}
The performance of the trained models has been measured based on their estimation error. Here, to enable a fair comparison with the previous study, the selected Figure of Merit is the \gls{MASE}. This metric measures a model's error as proportion of the naive estimator's error. For instance, a \gls{MASE} of $0.1$ would mean that the model's error is $10\%$ of the naive estimator's error. In this way, Figure \ref{fig:mase-best-kqi} shows the estimation error in terms of \gls{MASE} of each of the models \gls{SML} and \gls{STN}) as well as the best performing classical model for each of the \glspl{KQI} mentioned previously. The results obtained with each model are represented as functions of the number of input features.

\subsubsection{Latency}

Figure \ref{fig:mase-best-kqi}\subref{fig:best-latency} shows that the \gls{STN} model performs best when using a small number of features but becomes unstable as more features are added, resulting in fluctuating  resulting  \gls{MASE} values. This instability means that increasing the number of features does not necessarily improve accuracy. The \gls{STN} model achieves its best performance at 8 features, with a \gls{MASE} of $0.529$.

Conversely, the \gls{SML} model starts with the lowest performance when using a small number of features. However it  shows consistent improvement with increasing feature count, reaching its best performance at nine features, with a \gls{MASE} of $0.535$, before slightly degrading. 

Finally, \gls{SVR} demonstrates to have the most stable behavior across different feature configurations, though its lowest error, at 10 features, is slightly worse than both of the best \gls{STN} and \gls{SML} models, with a \gls{MASE} of $0.538$.

\subsubsection{Freeze Percentage}
The results in Figure \ref{fig:mase-best-kqi}\subref{fig:best-freeze} show that the \gls{RF} model performs best with a small number of features. However, its accuracy plateaus and is eventually surpassed by both \gls{SML} and \gls{STN} as more features are added. The best performance obtained with \gls{RF} is at 9 features, with a MASE of 0.139. Here, \gls{SML} demonstrates to have the most consistent performance, with a steady improvement up to nine features, where it obtains its best results at 0.135, and only a slight decline thereafter.

In contrast, the \gls{STN} model shows greater variability and only outperforms both, \gls{RF} and \gls{SML}, at 9 features, which is worth noting due to this being the number of features where they also exhibit their best performance; and at 13 features, where it achieves the lowest overall error, with a \gls{MASE} value of $0.119$. These results suggest that \textit{Freeze Percentage} prediction may benefit from models capable of capturing more complex feature interactions.

\begin{figure*}[!ht]
    \centering
    \includegraphics[width=\textwidth]{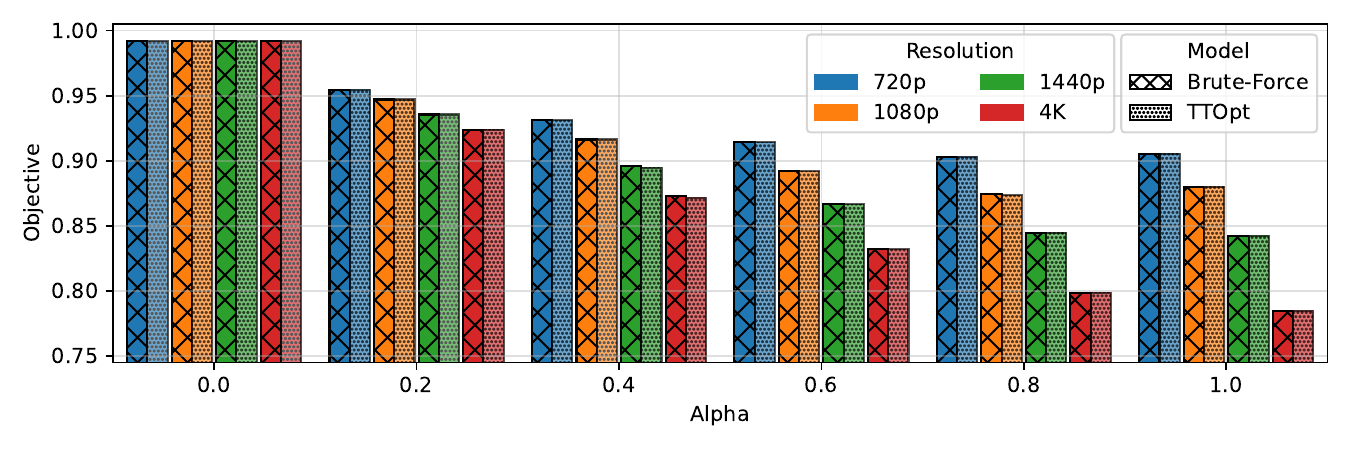}
    \vspace{-15pt}\caption{TTOpt vs Brute-force: Average objective values for different resolution requirements and alpha parameters.}
    \label{fig:ttopt-vs-bt-obj}
\end{figure*}
~
\begin{table*}[!ht]
\vspace{-5pt}
\caption{Results from 100 runs of the optimization procedures, objective values are visualized in Figure \ref{fig:ttopt-vs-bt-obj}. (*) Where $\alpha = 0$ all combinations of parameters with \glspl{PRB} = 5 will have the same objective function value, and the STN Optimizer finds many different combinations of these (5 is the minimum value of \glspl{PRB} that can be assigned to a user in our models).}
\label{tab:optimization-results}
\begin{tabularx}{\textwidth}{c ccccccc cccccc}
\toprule
 & \multicolumn{7}{c}{\textbf{Brute Force}} & \multicolumn{6}{c}{\textbf{STN Optimizer}} \\
\cmidrule(lr){3-8} \cmidrule(lr){9-14}

&  & \multicolumn{3}{c}{\shortstack{\textbf{Optimization} \\ \textbf{Parameters}}} & \multicolumn{2}{c}{\shortstack{\textbf{Predicted} \\ \textbf{KQIs}}} & \multirow{2}{*}{\shortstack{\textbf{Average} \\ \textbf{Time to} \\ \textbf{Solution} \\ \textbf{(ms)}} } & \multicolumn{3}{c}{\shortstack{\textbf{Optimization} \\ \textbf{Parameters}}} & \multicolumn{2}{c}{\shortstack{\textbf{Predicted} \\ \textbf{KQIs}}} & \multirow{2}{*}{\shortstack{\textbf{Average} \\ \textbf{Time to} \\ \textbf{Solution} \\ \textbf{(ms)} \\ \phantom{T}}}\\

\cmidrule(lr){3-5} \cmidrule(lr){6-7} \cmidrule(lr){9-11} \cmidrule(lr){12-13}

\textbf{Resolution} & $\boldsymbol{\alpha}$ & \textbf{PRBs} & \textbf{Resolution} & \textbf{FPS} & \shortstack{\textbf{Latency} \\ \textbf{(ms)}} & \textbf{EFPS} & & \textbf{PRBs} & \textbf{Resolution} & \textbf{FPS} & \shortstack{\textbf{Latency} \\ \textbf{(ms)}} & \textbf{EFPS} & \\
\midrule

\multirow{6}{*}{720p} & \multicolumn{1}{c|}{0} & \multicolumn{1}{c|}{5} & \multicolumn{1}{c|}{720p} & \multicolumn{1}{c|}{30} & \multicolumn{1}{c|}{127.58} & \multicolumn{1}{c|}{33} & 12.311 & \multicolumn{1}{c|}{5} & \multicolumn{1}{c|}{(*)} & \multicolumn{1}{c|}{(*)} & \multicolumn{1}{c|}{(*)} & \multicolumn{1}{c|}{(*)} & 10.542 \\ \cline{2-14} 
 & \multicolumn{1}{c|}{0.2} & \multicolumn{1}{c|}{13} & \multicolumn{1}{c|}{1440p} & \multicolumn{1}{c|}{120} & \multicolumn{1}{c|}{105.29} & \multicolumn{1}{c|}{118} & 12.892 & \multicolumn{1}{c|}{13} & \multicolumn{1}{c|}{1440p} & \multicolumn{1}{c|}{120} & \multicolumn{1}{c|}{105.29} & \multicolumn{1}{c|}{118} & 10.438 \\ \cline{2-14} 
 & \multicolumn{1}{c|}{0.4} & \multicolumn{1}{c|}{18} & \multicolumn{1}{c|}{1440p} & \multicolumn{1}{c|}{120} & \multicolumn{1}{c|}{89.62} & \multicolumn{1}{c|}{118} & 13.267 & \multicolumn{1}{c|}{18} & \multicolumn{1}{c|}{1440p} & \multicolumn{1}{c|}{120} & \multicolumn{1}{c|}{89.62} & \multicolumn{1}{c|}{118} & 10.884 \\ \cline{2-14} 
 & \multicolumn{1}{c|}{0.6} & \multicolumn{1}{c|}{22} & \multicolumn{1}{c|}{1440p} & \multicolumn{1}{c|}{120} & \multicolumn{1}{c|}{81.11} & \multicolumn{1}{c|}{118} & 12.530 & \multicolumn{1}{c|}{22} & \multicolumn{1}{c|}{1440p} & \multicolumn{1}{c|}{120} & \multicolumn{1}{c|}{81.11} & \multicolumn{1}{c|}{118} & 11.096 \\ \cline{2-14} 
 & \multicolumn{1}{c|}{0.8} & \multicolumn{1}{c|}{26} & \multicolumn{1}{c|}{1440p} & \multicolumn{1}{c|}{120} & \multicolumn{1}{c|}{75.45} & \multicolumn{1}{c|}{118} & 12.343 & \multicolumn{1}{c|}{26} & \multicolumn{1}{c|}{1440p} & \multicolumn{1}{c|}{120} & \multicolumn{1}{c|}{75.45} & \multicolumn{1}{c|}{118} & 11.232 \\ \cline{2-14} 
 & \multicolumn{1}{c|}{1} & \multicolumn{1}{c|}{91} & \multicolumn{1}{c|}{1440p} & \multicolumn{1}{c|}{120} & \multicolumn{1}{c|}{67.12} & \multicolumn{1}{c|}{116} & 12.364 & \multicolumn{1}{c|}{91} & \multicolumn{1}{c|}{1440p} & \multicolumn{1}{c|}{120} & \multicolumn{1}{c|}{67.12} & \multicolumn{1}{c|}{116} & 10.803 \\ 
 
 \midrule

\multirow{7}{*}{1080p} & \multicolumn{1}{c|}{0} & \multicolumn{1}{c|}{5} & \multicolumn{1}{c|}{720p} & \multicolumn{1}{c|}{30} & \multicolumn{1}{c|}{127.58} & \multicolumn{1}{c|}{33} & 12.318 & \multicolumn{1}{c|}{5} & \multicolumn{1}{c|}{(*)} & \multicolumn{1}{c|}{(*)} & \multicolumn{1}{c|}{(*)} & \multicolumn{1}{c|}{(*)} & 10.287 \\ \cline{2-14} 
 & \multicolumn{1}{c|}{0.2} & \multicolumn{1}{c|}{13} & \multicolumn{1}{c|}{1440p} & \multicolumn{1}{c|}{120} & \multicolumn{1}{c|}{105.29} & \multicolumn{1}{c|}{118} & 12.321 & \multicolumn{1}{c|}{13} & \multicolumn{1}{c|}{1440p} & \multicolumn{1}{c|}{120} & \multicolumn{1}{c|}{105.29} & \multicolumn{1}{c|}{118} & 10.487 \\ \cline{2-14} 
 & \multicolumn{1}{c|}{0.4} & \multicolumn{1}{c|}{18} & \multicolumn{1}{c|}{1440p} & \multicolumn{1}{c|}{120} & \multicolumn{1}{c|}{89.62} & \multicolumn{1}{c|}{118} & 12.395 & \multicolumn{1}{c|}{18} & \multicolumn{1}{c|}{1440p} & \multicolumn{1}{c|}{120} & \multicolumn{1}{c|}{89.62} & \multicolumn{1}{c|}{118} & 10.274 \\ \cline{2-14} 
 & \multicolumn{1}{c|}{0.6} & \multicolumn{1}{c|}{22} & \multicolumn{1}{c|}{1440p} & \multicolumn{1}{c|}{120} & \multicolumn{1}{c|}{81.11} & \multicolumn{1}{c|}{118} & 12.468 & \multicolumn{1}{c|}{22} & \multicolumn{1}{c|}{1440p} & \multicolumn{1}{c|}{120} & \multicolumn{1}{c|}{81.11} & \multicolumn{1}{c|}{118} & 10.505 \\ \cline{2-14} 
 & \multicolumn{1}{c|}{\multirow{2}{*}{0.8}} & \multicolumn{1}{c|}{\multirow{2}{*}{29}} & \multicolumn{1}{c|}{\multirow{2}{*}{4k}} & \multicolumn{1}{c|}{\multirow{2}{*}{120}} & \multicolumn{1}{c|}{\multirow{2}{*}{95.75}} & \multicolumn{1}{c|}{\multirow{2}{*}{125}} & \multirow{2}{*}{12.340} & \multicolumn{1}{c|}{26} & \multicolumn{1}{c|}{1440p} & \multicolumn{1}{c|}{120} & \multicolumn{1}{c|}{75.45} & \multicolumn{1}{c|}{118} & \multirow{2}{*}{10.515} \\
 & \multicolumn{1}{c|}{} & \multicolumn{1}{c|}{} & \multicolumn{1}{c|}{} & \multicolumn{1}{c|}{} & \multicolumn{1}{c|}{} & \multicolumn{1}{c|}{} &  & \multicolumn{1}{c|}{29} & \multicolumn{1}{c|}{4k} & \multicolumn{1}{c|}{120} & \multicolumn{1}{c|}{95.75} & \multicolumn{1}{c|}{125} &  \\ \cline{2-14} 
 & \multicolumn{1}{c|}{1} & \multicolumn{1}{c|}{91} & \multicolumn{1}{c|}{4k} & \multicolumn{1}{c|}{120} & \multicolumn{1}{c|}{86.04} & \multicolumn{1}{c|}{124} & 12.700 & \multicolumn{1}{c|}{91} & \multicolumn{1}{c|}{4k} & \multicolumn{1}{c|}{120} & \multicolumn{1}{c|}{86.04} & \multicolumn{1}{c|}{124} & 12.436 \\
 
 \midrule

\multirow{7}{*}{1440p} & \multicolumn{1}{c|}{0} & \multicolumn{1}{c|}{5} & \multicolumn{1}{c|}{720p} & \multicolumn{1}{c|}{30} & \multicolumn{1}{c|}{127.58} & \multicolumn{1}{c|}{33} & 12.293 & \multicolumn{1}{c|}{5} & \multicolumn{1}{c|}{(*)} & \multicolumn{1}{c|}{(*)} & \multicolumn{1}{c|}{(*)} & \multicolumn{1}{c|}{(*)} & 11.408 \\ \cline{2-14} 
 & \multicolumn{1}{c|}{0.2} & \multicolumn{1}{c|}{13} & \multicolumn{1}{c|}{1440p} & \multicolumn{1}{c|}{120} & \multicolumn{1}{c|}{105.29} & \multicolumn{1}{c|}{118} & 12.261 & \multicolumn{1}{c|}{13} & \multicolumn{1}{c|}{1440p} & \multicolumn{1}{c|}{120} & \multicolumn{1}{c|}{105.29} & \multicolumn{1}{c|}{118} & 10.641 \\ \cline{2-14} 
 & \multicolumn{1}{c|}{\multirow{2}{*}{0.4}} & \multicolumn{1}{c|}{\multirow{2}{*}{21}} & \multicolumn{1}{c|}{\multirow{2}{*}{4k}} & \multicolumn{1}{c|}{\multirow{2}{*}{120}} & \multicolumn{1}{c|}{\multirow{2}{*}{111.25}} & \multicolumn{1}{c|}{\multirow{2}{*}{125}} & \multirow{2}{*}{12.467} & \multicolumn{1}{c|}{18} & \multicolumn{1}{c|}{1440p} & \multicolumn{1}{c|}{120} & \multicolumn{1}{c|}{89.62} & \multicolumn{1}{c|}{118} & \multirow{2}{*}{10.355} \\
 & \multicolumn{1}{c|}{} & \multicolumn{1}{c|}{} & \multicolumn{1}{c|}{} & \multicolumn{1}{c|}{} & \multicolumn{1}{c|}{} & \multicolumn{1}{c|}{} &  & \multicolumn{1}{c|}{21} & \multicolumn{1}{c|}{4k} & \multicolumn{1}{c|}{120} & \multicolumn{1}{c|}{111.25} & \multicolumn{1}{c|}{125} &  \\ \cline{2-14} 
 & \multicolumn{1}{c|}{0.6} & \multicolumn{1}{c|}{25} & \multicolumn{1}{c|}{4k} & \multicolumn{1}{c|}{120} & \multicolumn{1}{c|}{101.72} & \multicolumn{1}{c|}{125} & 12.523 & \multicolumn{1}{c|}{25} & \multicolumn{1}{c|}{4k} & \multicolumn{1}{c|}{120} & \multicolumn{1}{c|}{101.72} & \multicolumn{1}{c|}{125} & 10.723 \\ \cline{2-14} 
 & \multicolumn{1}{c|}{0.8} & \multicolumn{1}{c|}{29} & \multicolumn{1}{c|}{4k} & \multicolumn{1}{c|}{120} & \multicolumn{1}{c|}{95.75} & \multicolumn{1}{c|}{125} & 12.677 & \multicolumn{1}{c|}{29} & \multicolumn{1}{c|}{4k} & \multicolumn{1}{c|}{120} & \multicolumn{1}{c|}{95.75} & \multicolumn{1}{c|}{125} & 10.525 \\ \cline{2-14} 
 & \multicolumn{1}{c|}{1} & \multicolumn{1}{c|}{91} & \multicolumn{1}{c|}{4k} & \multicolumn{1}{c|}{120} & \multicolumn{1}{c|}{86.043} & \multicolumn{1}{c|}{124} & 12.557 & \multicolumn{1}{c|}{91} & \multicolumn{1}{c|}{4k} & \multicolumn{1}{c|}{120} & \multicolumn{1}{c|}{86.043} & \multicolumn{1}{c|}{124} & 10.627 \\ 
 
\midrule

\multirow{7}{*}{4k} & \multicolumn{1}{c|}{0} & \multicolumn{1}{c|}{5} & \multicolumn{1}{c|}{720p} & \multicolumn{1}{c|}{30} & \multicolumn{1}{c|}{127.58} & \multicolumn{1}{c|}{33} & 12.433 & \multicolumn{1}{c|}{5} & \multicolumn{1}{c|}{(*)} & \multicolumn{1}{c|}{(*)} & \multicolumn{1}{c|}{(*)} & \multicolumn{1}{c|}{(*)} & 10.153 \\ \cline{2-14} 
 & \multicolumn{1}{c|}{0.2} & \multicolumn{1}{c|}{13} & \multicolumn{1}{c|}{1440p} & \multicolumn{1}{c|}{120} & \multicolumn{1}{c|}{105.29} & \multicolumn{1}{c|}{118} & 12.650 & \multicolumn{1}{c|}{13} & \multicolumn{1}{c|}{1440p} & \multicolumn{1}{c|}{120} & \multicolumn{1}{c|}{105.29} & \multicolumn{1}{c|}{118} & 10.619 \\ \cline{2-14} 
 & \multicolumn{1}{c|}{\multirow{2}{*}{0.4}} & \multicolumn{1}{c|}{\multirow{2}{*}{21}} & \multicolumn{1}{c|}{\multirow{2}{*}{4k}} & \multicolumn{1}{c|}{\multirow{2}{*}{120}} & \multicolumn{1}{c|}{\multirow{2}{*}{111.24}} & \multicolumn{1}{c|}{\multirow{2}{*}{124}} & \multirow{2}{*}{12.485} & \multicolumn{1}{c|}{18} & \multicolumn{1}{c|}{1440p} & \multicolumn{1}{c|}{120} & \multicolumn{1}{c|}{89.62} & \multicolumn{1}{c|}{118} & \multirow{2}{*}{10.542} \\
 & \multicolumn{1}{c|}{} & \multicolumn{1}{c|}{} & \multicolumn{1}{c|}{} & \multicolumn{1}{c|}{} & \multicolumn{1}{c|}{} & \multicolumn{1}{c|}{} &  & \multicolumn{1}{c|}{21} & \multicolumn{1}{c|}{4k} & \multicolumn{1}{c|}{120} & \multicolumn{1}{c|}{111.24} & \multicolumn{1}{c|}{124} &  \\ \cline{2-14} 
 & \multicolumn{1}{c|}{0.6} & \multicolumn{1}{c|}{25} & \multicolumn{1}{c|}{4k} & \multicolumn{1}{c|}{120} & \multicolumn{1}{c|}{101.72} & \multicolumn{1}{c|}{124} & 12.645 & \multicolumn{1}{c|}{25} & \multicolumn{1}{c|}{4k} & \multicolumn{1}{c|}{120} & \multicolumn{1}{c|}{101.72} & \multicolumn{1}{c|}{124} & 10.670 \\ \cline{2-14} 
 & \multicolumn{1}{c|}{0.8} & \multicolumn{1}{c|}{29} & \multicolumn{1}{c|}{4k} & \multicolumn{1}{c|}{120} & \multicolumn{1}{c|}{95.75} & \multicolumn{1}{c|}{125} & 12.633 & \multicolumn{1}{c|}{29} & \multicolumn{1}{c|}{4k} & \multicolumn{1}{c|}{120} & \multicolumn{1}{c|}{95.75} & \multicolumn{1}{c|}{125} & 10.453 \\ \cline{2-14} 
 & \multicolumn{1}{c|}{1} & \multicolumn{1}{c|}{91} & \multicolumn{1}{c|}{4k} & \multicolumn{1}{c|}{120} & \multicolumn{1}{c|}{86.04} & \multicolumn{1}{c|}{124} & 12.739 & \multicolumn{1}{c|}{91} & \multicolumn{1}{c|}{4k} & \multicolumn{1}{c|}{120} & \multicolumn{1}{c|}{86.04} & \multicolumn{1}{c|}{124} & 10.261 \\
 \bottomrule
\end{tabularx}
\end{table*}

\subsubsection{Effective Frames per Second}
Figure \ref{fig:mase-best-kqi}\subref{fig:best-efps} shows that the classical model (\gls{KNR}) performs well with a small number of features but its accuracy deteriorates as more features are added, achieving its best \gls{MASE} of $0.084$, at 5 features. Similarly to the previous \gls{KQI}, the \gls{SML} model demonstrates stable and consistent improvements as the number of features increases, ultimately surpassing \gls{KNR} and achieving the best overall performance, with a \gls{MASE} score of $0.076$ at 13 features.

For this last \gls{KQI}, the \gls{STN} model yields comparatively lower accuracy, exhibiting the highest error across all feature configurations. It achieves its best performance at 10 features, with a \gls{MASE} of $0.109$, which presents a large gap to the results obtained with the other models. However, the steady decrease of error for the \gls{SML} estimator increases might be indicative of the method's ability to extract useful information from the less relevant metrics to improve the estimation, meaning that including additional metrics could further improve the estimation of the \textit{Effective Frames Per Second} \gls{KQI}.

\subsection{Model Loading and Inference Time}

In addition to estimation accuracy, the model loading and inference times were evaluated for all models, averaging over 100 runs on a test set of 694 entries. This evaluation was carried out in a docker instance on an Apple M1 Pro Macbook, with 8GB RAM provided to the container. 
Furthermore, for simplicity, only the models with 13 features were evaluated for each \gls{KQI}.
Here, the classical models, which are used as a benchmark, showed consistent results across \glspl{KQI}: \gls{SVR} (Latency) requires 7.9 ms to load the model, and completes the inference on the test set in 8.7 ms; \gls{RF} (Freeze Percentage) loads in 8 ms, with the inference lasting 8.7 ms as well;  \gls{KNR} (EFPS) loads in 7.9 ms, and has an inference time of 10 ms.

Moving on to the \gls{SML} models, they consistently perform well in these metrics. For Latency the model loads in just 0.7 ms and performs inference in 7.5 ms. For both Freeze Percent and EFPS the models take only 0.5 ms to load, while achieving inference times of 6.1 ms and 9.4 ms respectively. This shows the advantage of using \gls{QC} for training, as the \gls{SML} model are smaller than the fully classical benchmark, achieving sub millisecond model loading times.

In contrast to \gls{SML}, the \gls{STN} models come as the slowest of all models tried, with the Latency model loading in 13.2 ms, and completing the inference in 68.6 ms, the Freeze Percentage model taking 12.9 ms and 11.3 ms to load and infer, respectively, and the EFPS model loading in 15.1 ms, and completing the inference in 17.7 ms. The fact that the \gls{STN} models are slower in terms of loading times comes as no surprise, since they attempt to model more complex, and higher order feature interactions. The increase in inference time can also be explain due to the fact that, for each point in the test set used, the models have to perform tensor contractions, which are more complex operations in relation to the ones used by the classical models analyzed.

\subsection{Optimization Evaluation}

To evaluate the \gls{STN} algorithm for the network parameter optimization use case, the proposed method was benchmarked against a brute-force search under various configurations. Here, the goal is to identify the optimal allocation of \gls{PRB}s, resolution settings, and frame rates that maximize user \gls {QoE} while minimizing network resource consumption.

The results of this evaluation are shown in Figure \ref{fig:ttopt-vs-bt-obj}, and the performance data is provided in Table \ref{tab:optimization-results}. These present the optimization outcomes across a range of trade-off values between the service and network costs (denoted by the parameter $\alpha$) and resolution constraints.

As shown in figure \ref{fig:ttopt-vs-bt-obj}, the \gls{STN} algorithm adapts its performance according to the value of $\alpha$. Higher values of $\alpha$ emphasize user experience, while lower values prioritize network resource efficiency. The algorithm was executed 100 times for each configuration, and the plot displays the average results across these runs.

The results demonstrate that \gls{STN} effectively finds the optimal configuration for this problem. The actual optimum value and the one found by \gls{STN} match in most cases. However, a difference of around $0.15\%$ in the objective function's value is observed at a few points on the graph. This small deviation suggests that \gls{STN} provides near-optimal solutions while significantly reducing the number of function evaluations, making it a promising candidate for scaling to higher-dimensional problems. 

Table \ref{tab:optimization-results} provides a deeper look into the optimization results, listing the optimal parameters identified for each combination of $\alpha$ and resolution. It includes the predicted \gls{KQI}s—Latency and Effective FPS—as well as the average time taken to find each solution. The \gls{STN} optimizer achieves these results with considerably lower computational effort. For instance, at $\alpha = 0.8$ and a minimum resolution of 1440p, both brute-force and \gls{STN} arrive at the same optimal configuration (29 PRBs, 4K resolution, 120 FPS), but \gls{STN} does so in just 10.5 ms, compared to 12.6 ms for brute-force.

Interestingly, when $\alpha = 0$, indicating a focus solely on minimizing resource usage, the \gls{STN} optimizer identifies several equally valid configurations, as all parameter sets with the minimum PRB value yield identical objective scores. These cases are marked with asterisks (*) in Table \ref{tab:optimization-results}, highlighting \gls{STN}’s adaptability in exploring flat or ambiguous solution spaces.

\section{Conclusions}\label{sec:conclusions}

\gls{QML} and \gls{QI} algorithms can be useful for boosting cellular network management processes due to their outstanding scalability and ability to solve complex problems that classical \gls{ML} may struggle with. The present article also proposed the integration of this technology into the cellular network management system using the \gls{ORAN} architecture. To validate the usefulness of \gls{QML} and \gls{QI} for management, a case study for optimizing the \gls{QoE} of \gls{CG} was proposed. For this optimization, first, an estimator is trained to obtain \gls{QoE} from network metrics and service configuration that, then, is used by the optimizer to find the optimal solution. 

Regarding the estimator, two methods were proposed, \gls{SML} and \gls{STN}, and were compared to the best, reproducible estimator from a previous work. The results showed that the quantum-based approaches obtained lower estimation errors and they were able to achieve a steady betterment of the estimation as the number of features increased, which could be indicative that these approaches were able to extract useful information from more complex interactions between the features.

In regards to the optimizer, \gls{STN} was used to find the optimum solutions for the network and service configurations using an objective function to balance resource usage and perceived quality. Then, the algorithm was compared against a brute-force search and the results showed that the optimizer was able to achieve the optimum solution for most cases, with only a few configurations showing a marginal decrease in the objective value, while reducing the time to solution around a $15\%$.

The results showed that the \gls{QML} and \gls{QI} algorithms were able to outperform all the classical methods, however, it is noteworthy to point out the limitations of the data used for this work. Although the quantum-based approaches improved compared to previous methods, the setup of the testbed were data was obtained and the amount of different configurations limits the complexity of the dataset, with classical methods also obtaining great results. For this reason, it would be interesting to extend further the analysis and assessment of capabilities of these approaches against much more complex datasets were classical \gls{ML} struggles.

\bibliography{Bib}

\begin{thebibliography}{10}
\providecommand{\url}[1]{#1}
\csname url@samestyle\endcsname
\providecommand{\newblock}{\relax}
\providecommand{\bibinfo}[2]{#2}
\providecommand{\BIBentrySTDinterwordspacing}{\spaceskip=0pt\relax}
\providecommand{\BIBentryALTinterwordstretchfactor}{4}
\providecommand{\BIBentryALTinterwordspacing}{\spaceskip=\fontdimen2\font plus
\BIBentryALTinterwordstretchfactor\fontdimen3\font minus
  \fontdimen4\font\relax}
\providecommand{\BIBforeignlanguage}[2]{{%
\expandafter\ifx\csname l@#1\endcsname\relax
\typeout{** WARNING: IEEEtran.bst: No hyphenation pattern has been}%
\typeout{** loaded for the language `#1'. Using the pattern for}%
\typeout{** the default language instead.}%
\else
\language=\csname l@#1\endcsname
\fi
#2}}
\providecommand{\BIBdecl}{\relax}
\BIBdecl

\bibitem{Sundsvist2023Bottleneck}
T.~Sundqvist, M.~Bhuyan, and E.~Elmroth, ``{Bottleneck identification and
  failure prevention with procedural learning in 5G RAN},'' in \emph{{2023
  IEEE/ACM 23rd International Symposium on Cluster, Cloud and Internet
  Computing (CCGrid)}}, 2023, pp. 426--436.

\bibitem{Kawasaki2023eBPFPrediction}
J.~Kawasaki, D.~Koyama, T.~Miyasaka, and T.~Otani, ``{Failure Prediction in
  Cloud Native 5G Core With eBPF-based Observability},'' in \emph{{2023 IEEE
  97th Vehicular Technology Conference (VTC2023-Spring)}}, 2023, pp. 1--6.

\bibitem{Dey2025HOV2X}
M.~R. Dey and M.~Patra, ``{Secure Handover in 5G-V2X: Detecting and Localizing
  DoS Attack to Ensure Reliable Communication},'' \emph{IEEE Transactions on
  Intelligent Transportation Systems}, pp. 1--14, 2025.

\bibitem{Zhao2024ResAllocRL}
Z.~Di, Z.~Zhong, Q.~Pengfei, Q.~Hao, and S.~Bin, ``{Resource allocation in
  multi-user cellular networks: A transformer-based deep reinforcement learning
  approach},'' \emph{China Communications}, vol.~21, no.~5, pp. 77--96, 2024.

\bibitem{Zhang2024ResAllocQoE}
L.~Zhang, X.~Zhou, J.~Zou, and X.~Liao, ``{A Resource Allocation Method for
  Multi-Cell Cellular Networks with System Capacity and QoE Tradeoff},'' in
  \emph{{2024 4th International Conference on Electronic Information
  Engineering and Computer Science (EIECS)}}, 2024, pp. 1087--1092.

\bibitem{Orús2019}
\BIBentryALTinterwordspacing
R.~Or{\'u}s, ``{Tensor networks for complex quantum systems},'' \emph{Nature
  Reviews Physics}, vol.~1, no.~9, pp. 538--550, Sep 2019. [Online]. Available:
  \url{https://doi.org/10.1038/s42254-019-0086-7}
\BIBentrySTDinterwordspacing

\bibitem{Nielsen_Chuang_2010}
M.~A. Nielsen and I.~L. Chuang, \emph{{Quantum Computation and Quantum
  Information: 10th Anniversary Edition}}.\hskip 1em plus 0.5em minus
  0.4em\relax Cambridge University Press, 2010.

\bibitem{Preskill_2018}
\BIBentryALTinterwordspacing
J.~Preskill, ``Quantum {C}omputing in the {NISQ} era and beyond,''
  \emph{{Quantum}}, vol.~2, p.~79, Aug. 2018. [Online]. Available:
  \url{https://doi.org/10.22331/q-2018-08-06-79}
\BIBentrySTDinterwordspacing

\bibitem{RevModPhys.90.015002}
\BIBentryALTinterwordspacing
T.~Albash and D.~A. Lidar, ``Adiabatic quantum computation,'' \emph{Rev. Mod.
  Phys.}, vol.~90, p. 015002, Jan 2018. [Online]. Available:
  \url{https://link.aps.org/doi/10.1103/RevModPhys.90.015002}
\BIBentrySTDinterwordspacing

\bibitem{10.3389/fphy.2014.00005}
\BIBentryALTinterwordspacing
A.~Lucas, ``{Ising formulations of many NP problems},'' \emph{Frontiers in
  Physics}, vol.~2, 2014. [Online]. Available:
  \url{https://www.frontiersin.org/journals/physics/articles/10.3389/fphy.2014.00005}
\BIBentrySTDinterwordspacing

\bibitem{Biamonte2017}
\BIBentryALTinterwordspacing
J.~Biamonte, P.~Wittek, N.~Pancotti, P.~Rebentrost, N.~Wiebe, and S.~Lloyd,
  ``{Quantum machine learning},'' \emph{Nature}, vol. 549, no. 7671, pp.
  195--202, Sep 2017. [Online]. Available:
  \url{https://doi.org/10.1038/nature23474}
\BIBentrySTDinterwordspacing

\bibitem{PhysRevA.101.032308}
\BIBentryALTinterwordspacing
M.~Schuld, A.~Bocharov, K.~M. Svore, and N.~Wiebe, ``{Circuit-centric quantum
  classifiers},'' \emph{Phys. Rev. A}, vol. 101, p. 032308, Mar 2020. [Online].
  Available: \url{https://link.aps.org/doi/10.1103/PhysRevA.101.032308}
\BIBentrySTDinterwordspacing

\bibitem{Sengupta2022-yd}
\BIBentryALTinterwordspacing
R.~Sengupta, S.~Adhikary, I.~Oseledets, and J.~Biamonte, ``"tensor networks in
  machine learning",'' \emph{Eur. Math. Soc. Mag.}, no. 126, pp. 4--12, Oct.
  2022. [Online]. Available:
  \url{https://euromathsoc.org/magazine/articles/101}
\BIBentrySTDinterwordspacing

\bibitem{Hildebrad2023QuantumComms}
B.~Hildebrand, A.~Ghimire, F.~Amsaad, A.~Razaque, and S.~P. Mohanty, ``Quantum
  communication networks: Design, reliability, and security,'' \emph{IEEE
  Potentials}, pp. 2--8, 2023.

\bibitem{BENNETT20147QKD}
\BIBentryALTinterwordspacing
C.~H. Bennett and G.~Brassard, ``{Quantum cryptography: Public key distribution
  and coin tossing},'' \emph{Theoretical Computer Science}, vol. 560, pp.
  7--11, 2014, theoretical Aspects of Quantum Cryptography – celebrating 30
  years of BB84. [Online]. Available:
  \url{https://www.sciencedirect.com/science/article/pii/S0304397514004241}
\BIBentrySTDinterwordspacing

\bibitem{Srikar2023QuantumBBU}
S.~Kasi, P.~Warburton, J.~Kaewell, and K.~Jamieson, ``{},'' \emph{IEEE
  Transactions on Quantum Engineering}, vol.~4, pp. 1--17, 2023.

\bibitem{LeTung2025QGNN}
L.~T. Giang, N.~X. Tung, and W.-J. Hwang, ``{Quantum Graph Neural Network for
  Resource Management in Wireless Communication},'' in \emph{{2025
  International Conference on Artificial Intelligence in Information and
  Communication (ICAIIC)}}, 2025, pp. 0128--0130.

\bibitem{QuantumORANMagazine}
\BIBentryALTinterwordspacing
S.~Fortes, J.~Villegas, F.~Chaouech, A.~Pereira, A.~Villarino, A.~Cortines,
  C.~Baena, R.~Orús, and R.~Barco, ``{Quantum-based Intelligence for Cellular
  Network Management: Use Cases, Challenges, and O-RAN Integration},'' Jun.
  2025. [Online]. Available:
  \url{http://dx.doi.org/10.36227/techrxiv.174952955.52927388/v1}
\BIBentrySTDinterwordspacing

\bibitem{EricssonQuantum}
A.~J. Awan, M.~A. Ullah, C.~J. Yang, R.~P. Sircar, B.~Grafulla-Gonz\'alez, and
  C.~Granbom', ``{Exploring the potential advantages of quantum computing in
  telecommunication networks},'' \emph{{Ericsson Technology Review}}, 2025.

\bibitem{KQIModeling}
A.~Herrera-Garcia, S.~Fortes, E.~Baena, J.~Mendoza, C.~Baena, and R.~Barco,
  ``{Modeling of Key Quality Indicators for End-to-End Network Management:
  Preparing for 5G},'' \emph{IEEE Vehicular Technology Magazine}, vol.~14,
  no.~4, pp. 76--84, 2019.

\bibitem{k0w8-qz67-22}
\BIBentryALTinterwordspacing
C.~Baena, O.~S. Peñaherrera-Pulla, L.~Camacho, R.~Barco, and S.~Fortes, ``{E2E
  dataset of Video Streaming and Cloud Gaming services over 4G and 5G},''
  \emph{IEEE Dataport}, 2022. [Online]. Available:
  \url{https://dx.doi.org/10.21227/k0w8-qz67}
\BIBentrySTDinterwordspacing

\bibitem{10148946}
C.~Baena, O.~S. Pe{\~{n}}aherrera-Pulla, L.~Camacho, R.~Barco, and S.~Fortes,
  ``{Video Streaming and Cloud Gaming Services Over 4G and 5G: A Complete
  Network and Service Metrics Dataset},'' \emph{IEEE Communications Magazine},
  vol.~61, no.~9, pp. 154--160, 2023.

\bibitem{InformationTheory}
\BIBentryALTinterwordspacing
A.~Orlitsky, ``{Information Theory},'' in \emph{{Encyclopedia of Physical
  Science and Technology (Third Edition)}}, 3rd~ed., R.~A. Meyers, Ed.\hskip
  1em plus 0.5em minus 0.4em\relax New York: Academic Press, 2003, pp.
  751--769. [Online]. Available:
  \url{https://www.sciencedirect.com/science/article/pii/B0122274105003379}
\BIBentrySTDinterwordspacing

\bibitem{pmlr-v25-neven12}
\BIBentryALTinterwordspacing
H.~Neven, V.~S. Denchev, G.~Rose, and W.~G. Macready, ``{QBoost: Large Scale
  Classifier Training with Adiabatic Quantum Optimization},'' in
  \emph{{Proceedings of the Asian Conference on Machine Learning}}, ser.
  Proceedings of Machine Learning Research, S.~C.~H. Hoi and W.~Buntine, Eds.,
  vol.~25.\hskip 1em plus 0.5em minus 0.4em\relax Singapore Management
  University, Singapore: PMLR, 04--06 Nov 2012, pp. 333--348. [Online].
  Available: \url{https://proceedings.mlr.press/v25/neven12.html}
\BIBentrySTDinterwordspacing

\bibitem{neven2009traininglargescaleclassifier}
\BIBentryALTinterwordspacing
{H. Neven, V. S. Denchev, G. Rose, and W. G. Macready}, ``{Training a Large
  Scale Classifier with the Quantum Adiabatic Algorithm},'' 2009. [Online].
  Available: \url{https://arxiv.org/abs/0912.0779}
\BIBentrySTDinterwordspacing

\bibitem{glover2019tutorialformulatingusingqubo}
\BIBentryALTinterwordspacing
F.~Glover, G.~Kochenberger, and Y.~Du, ``A tutorial on formulating and using
  qubo models,'' 2019. [Online]. Available:
  \url{https://arxiv.org/abs/1811.11538}
\BIBentrySTDinterwordspacing

\bibitem{góes2021qboostregressionproblemssolving}
\BIBentryALTinterwordspacing
C.~B.~D. Góes, T.~O. Maciel, G.~G. Pollachini, R.~Cuenca, J.~P. L.~C. Salazar,
  and E.~I. Duzzioni, ``{QBoost for regression problems: solving partial
  differential equations},'' 2021. [Online]. Available:
  \url{https://arxiv.org/abs/2108.13346}
\BIBentrySTDinterwordspacing

\bibitem{rogers2019floatingpointcalculationsquantumannealer}
\BIBentryALTinterwordspacing
M.~L. Rogers and R.~L.~S. Jr, ``{Floating-Point Calculations on a Quantum
  Annealer: Division and Matrix Inversion},'' 2019. [Online]. Available:
  \url{https://arxiv.org/abs/1901.06526}
\BIBentrySTDinterwordspacing

\bibitem{novikov2017exponentialmachines}
\BIBentryALTinterwordspacing
A.~Novikov, M.~Trofimov, and I.~Oseledets, ``{Exponential Machines},'' 2017.
  [Online]. Available: \url{https://arxiv.org/abs/1605.03795}
\BIBentrySTDinterwordspacing

\bibitem{TTOpt}
\BIBentryALTinterwordspacing
K.~Sozykin, A.~Chertkov, R.~Schutski, A.-H. Phan, A.~S. CICHOCKI, and
  I.~Oseledets, ``{TTOpt: A Maximum Volume Quantized Tensor Train-based
  Optimization and its Application to Reinforcement Learning},'' in
  \emph{{Advances in Neural Information Processing Systems}}, S.~Koyejo,
  S.~Mohamed, A.~Agarwal, D.~Belgrave, K.~Cho, and A.~Oh, Eds., vol.~35.\hskip
  1em plus 0.5em minus 0.4em\relax Curran Associates, Inc., 2022, pp.
  26\,052--26\,065. [Online]. Available:
  \url{https://proceedings.neurips.cc/paper_files/paper/2022/file/a730abbcd6cf4a371ca9545db5922442-Paper-Conference.pdf}
\BIBentrySTDinterwordspacing

\bibitem{Goreinov2001}
S.~Goreinov and E.~Tyrtyshnikov, ``{The maximal-volume concept in approximation
  by low-rank matrices},'' \emph{Contemporary Mathematics}, vol. 208, 01 2001.

\bibitem{amarisoft}
``{Amarisoft},'' \url{https://www.amarisoft.com/technology/}, visited in
  February 2024.

\bibitem{Baena2021}
C.~Baena, S.~Fortes, O.~Peñaherrera, and R.~Barco, ``{A Framework to boost the
  potential of network-in-a-box solutions},'' in \emph{2021 12th International
  Conference on Network of the Future (NoF)}, 2021, pp. 1--3.

\bibitem{IBMSML}
\BIBentryALTinterwordspacing
{IBM Quantum}, ``Multiverse computing - singularity integration guide,'' 2024,
  accessed: 2025-06-01. [Online]. Available:
  \url{https://docs.quantum.ibm.com/guides/multiverse-computing-singularity}
\BIBentrySTDinterwordspacing

\bibitem{BAENA2023109808}
\BIBentryALTinterwordspacing
C.~Baena, O.~Peñaherrera-Pulla, R.~Barco, and S.~Fortes, ``{Measuring and
  estimating Key Quality Indicators in Cloud Gaming services},'' \emph{Computer
  Networks}, vol. 231, p. 109808, 2023. [Online]. Available:
  \url{https://www.sciencedirect.com/science/article/pii/S1389128623002530}
\BIBentrySTDinterwordspacing

\end{thebibliography}
\bibliographystyle{IEEEtran}

\newcommand{\biowidth}{1in}

\begin{IEEEbiography}[{\includegraphics[width=\biowidth,trim={0.7in 0in 0.7in 0in},clip]{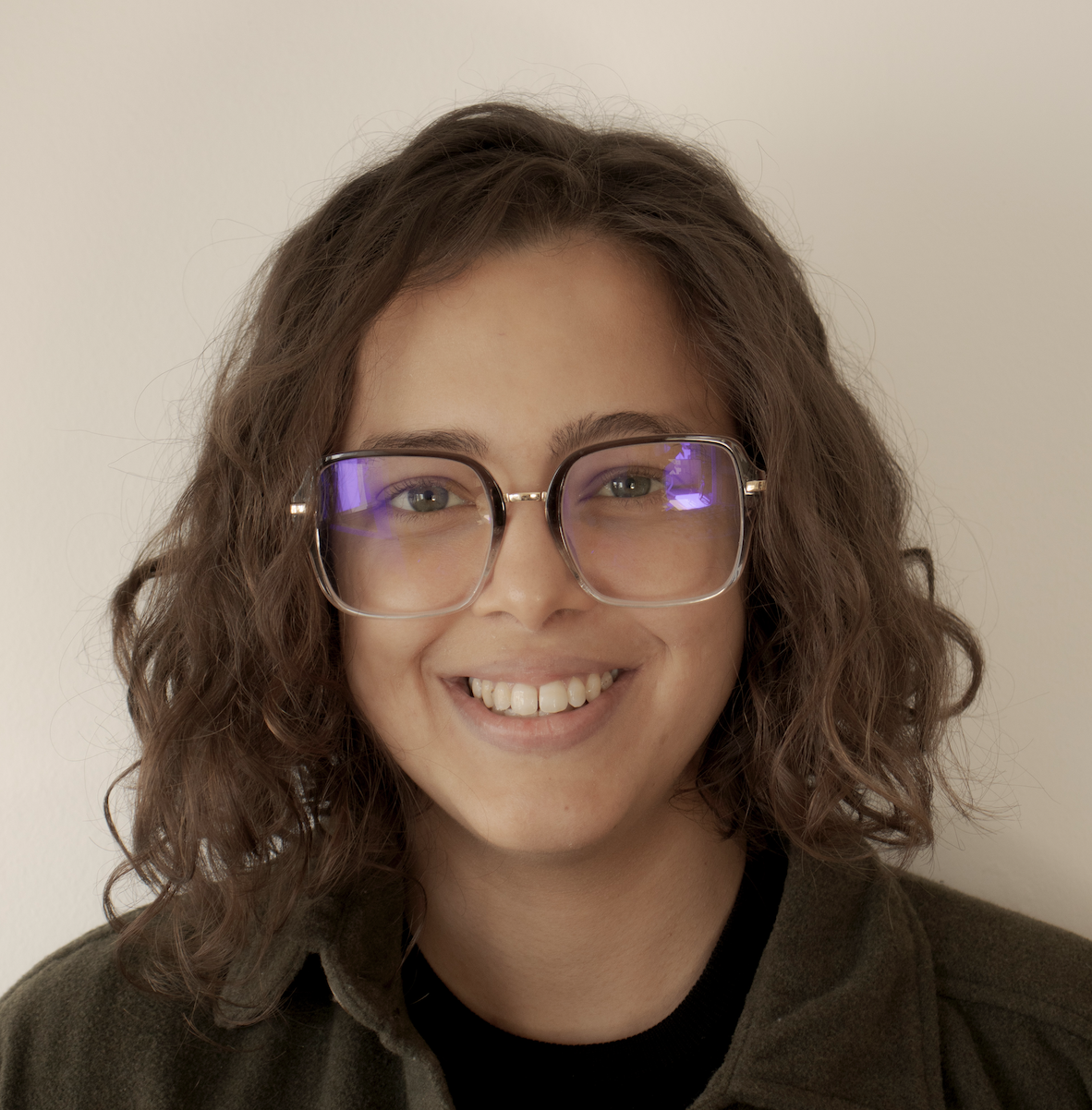}}]
{Fatma Chaouech}
\textit{ (fatma.chaouech@multiverse computing.com)}
received her M.Sc. in Software Engineering from the National Institute of Applied Science and Technology of Tunisia. In 2024, she joined Multiverse Computing as a Machine Learning Engineer, contributing to the development of classical and quantum ML libraries and applying them to solve use cases in telecommunications and healthcare.
\end{IEEEbiography}

\begin{IEEEbiography}[{\includegraphics[width=\biowidth,trim=0.2in 0in 0.4in 0in,clip]{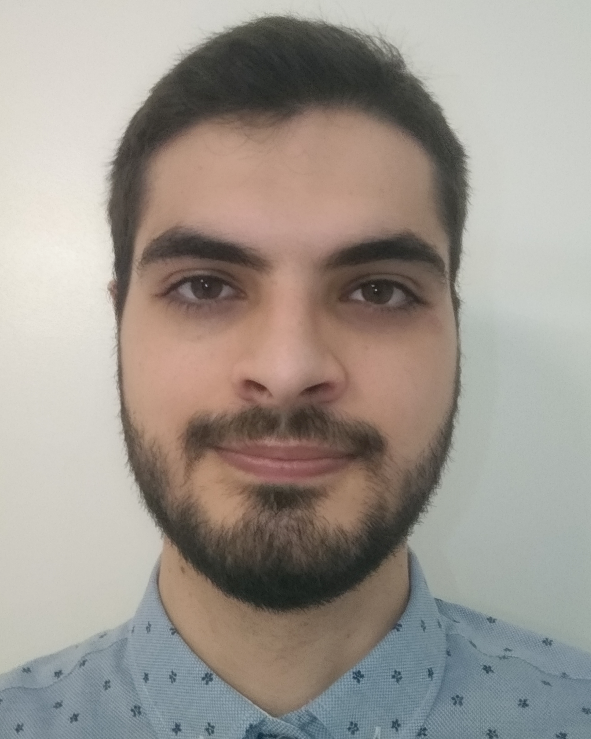}}]
{Javier Villegas}
\textit{ (jvc@ic.uma.es)}
received his degree in Telecommunications Systems Engineering and his M.Sc. degrees in telecommunication engineering and in telematic engineering from the University of M\'alaga, Spain. Currently, he is working as a Assistant Professor with the Department of Communications Engineering at the University of M\'alaga, where he is pursuing a Ph.D. 
\end{IEEEbiography}

\begin{IEEEbiography}[{\includegraphics[width=\biowidth,trim=0.9in 0in 0.9in 0in,clip]{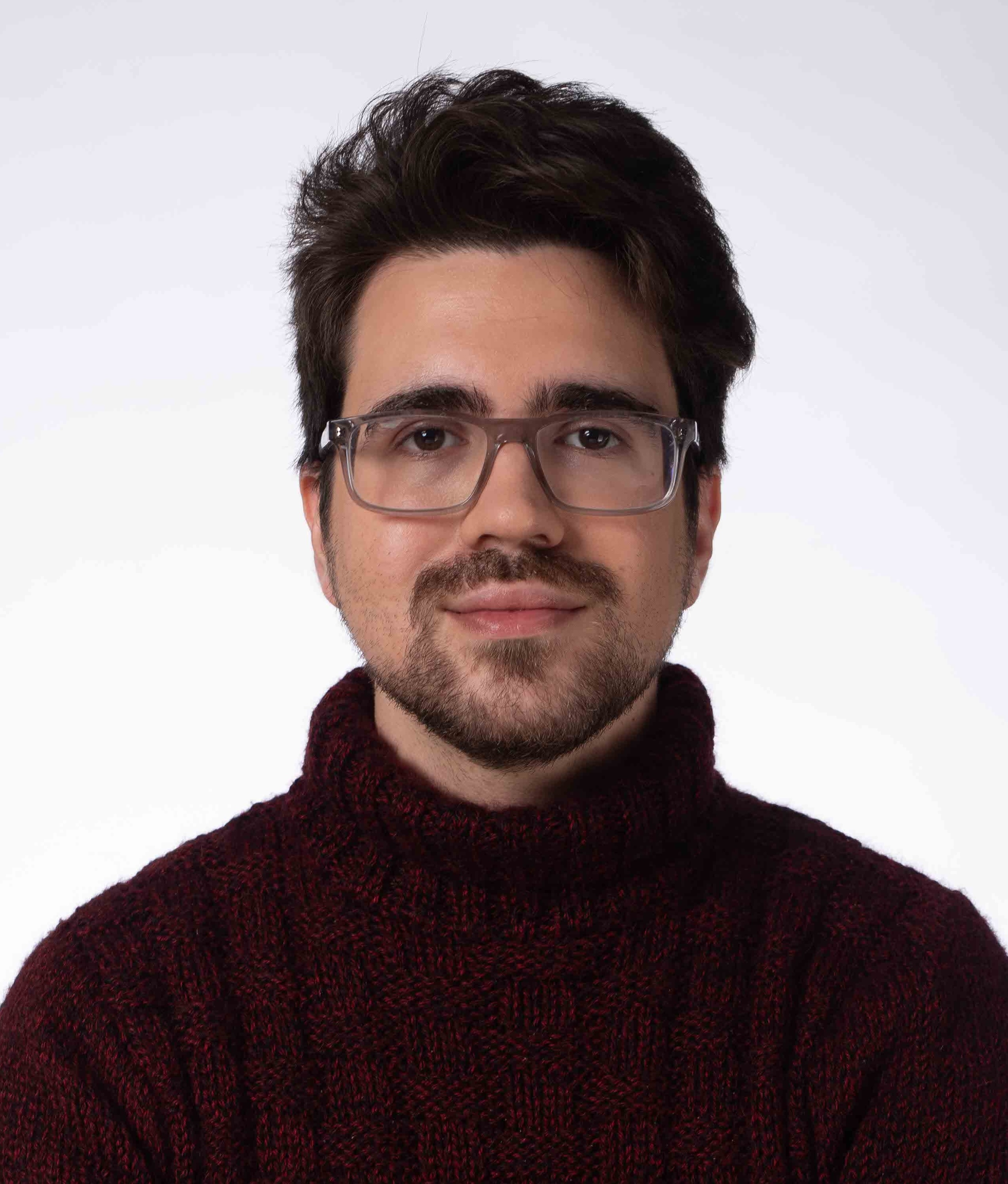}}]
{António Pereira}
\textit{ (antonio.pereira@multiverse computing.com)}
is an Integrated Master’s (B.Sc. + M.Sc.) student in Physics Engineering from the University of Minho. In 2023, he joined Multiverse Computing as a Quantum Software Engineer, where he developed expertise in quantum algorithms and tensor networks. His experience includes work on quantum and quantum-inspired solutions for finance.
\end{IEEEbiography}

\begin{IEEEbiography}[{\includegraphics[width=\biowidth,trim=0in 0in 0in 0in,clip]{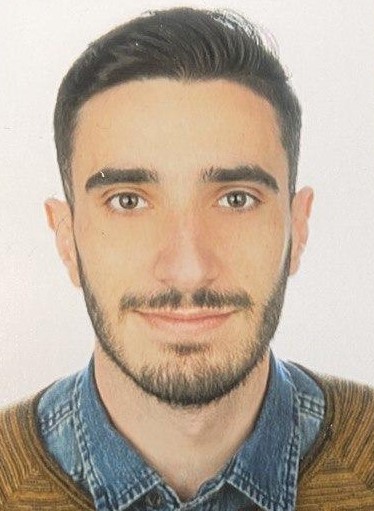}}]
{Carlos Baena}
\textit{ (jcbg@ic.uma.es)}
obtained his Ph.D. in Mobile Networks from the University of Málaga, Spain. His research specializes in the optimization of end-to-end (E2E) network performance through service-based approaches. He focuses on enhancing the user experience , particularly in applications related to video streaming and gaming, by leveraging machine learning (ML) techniques and key quality indicators (KQI) to optimize network resource management and overall performance.
\end{IEEEbiography}

\begin{IEEEbiography}[{\includegraphics[width=\biowidth,trim=0.2in 0in 0.2in 0in,clip]{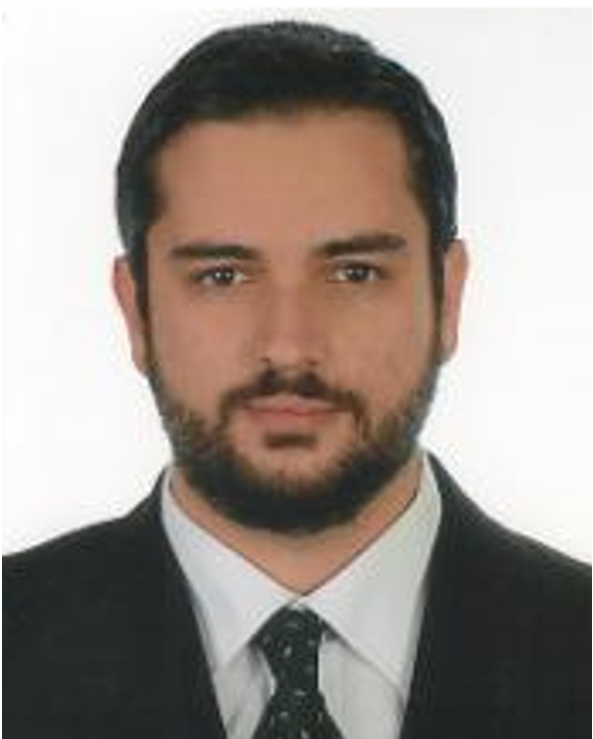}}]
     {Sergio Fortes}
     \textit{ (sfr@ic.uma.es, corresponding author)}
     is Associate Professor at the University of M\'alaga, from which it holds a M.Sc. and a Ph.D. in Telecommunication Engineering. He began his career being part of main european space agencies (DLR, CNES, ESA) and Avanti Communications plc, where he participated in various research and consultant activities on broadband and aeronautical satellite communications. In 2012, he joined the University of M\'alaga, where his topics of interest include cellular communications, satellite systems, smart-city, self-organizing / zero-touch networks (SON/ZSN), cloud robotics, and advanced applications of AI and machine learning techniques.  
\end{IEEEbiography}

\begin{IEEEbiography}[{\includegraphics[width=\biowidth,trim=0.2in 0in 0.2in 0in,clip]{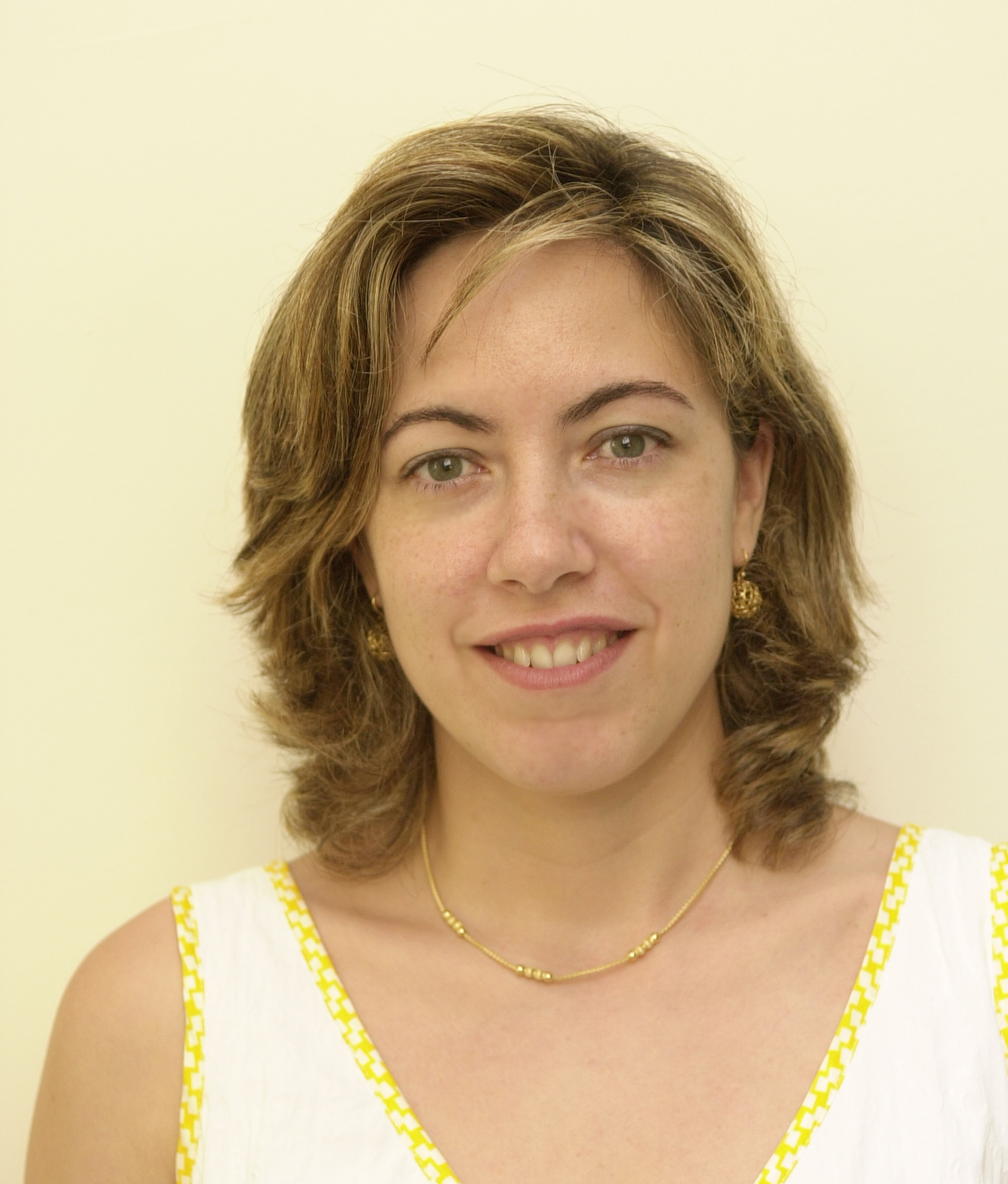}}]
{Raquel Barco}
\textit{ (rbm@ic.uma.es)}
received the M.Sc. and Ph.D. degrees in telecommunications from the University of Málaga. She has worked at Telefónica, Spain, and European Space Agency. She is currently a Full Professor with the University of Málaga. She participated in the Mobile Communication Systems Competence Center, jointly created by Nokia and UMA. She has published more than 120 scientific papers, filed several patents, and has lead projects with major companies.
\end{IEEEbiography}

\begin{IEEEbiography}[{\includegraphics[width=\biowidth,trim=0.2in 0in 0.2in 0in,clip]{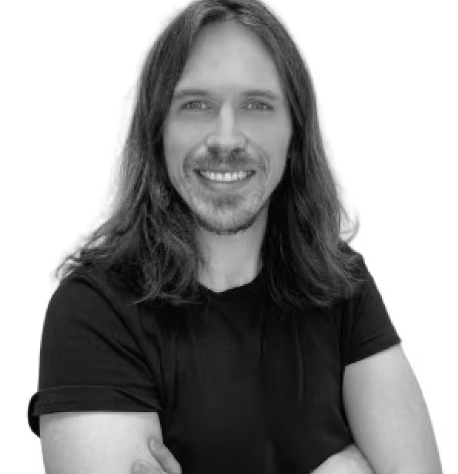}}]
{Dominic Gribben}
\textit{(dominic.gribben@multiverse
computing.com)}
holds an MSci in Physics with Theoretical Physics from the University of Nottingham and a PhD in Physics from the University of St Andrews. After completing a postdoctoral position at Johannes Gutenberg University Mainz he joined Multiverse computing in 2024 as a Tensor Network Engineer. In this role he has drawn on his academic background to apply quantum-inspired approaches to various practical problems, particularly in finance.
\end{IEEEbiography}

\begin{IEEEbiography}[{\includegraphics[width=\biowidth,trim=0.2in 0in 0.2in 0in,clip]{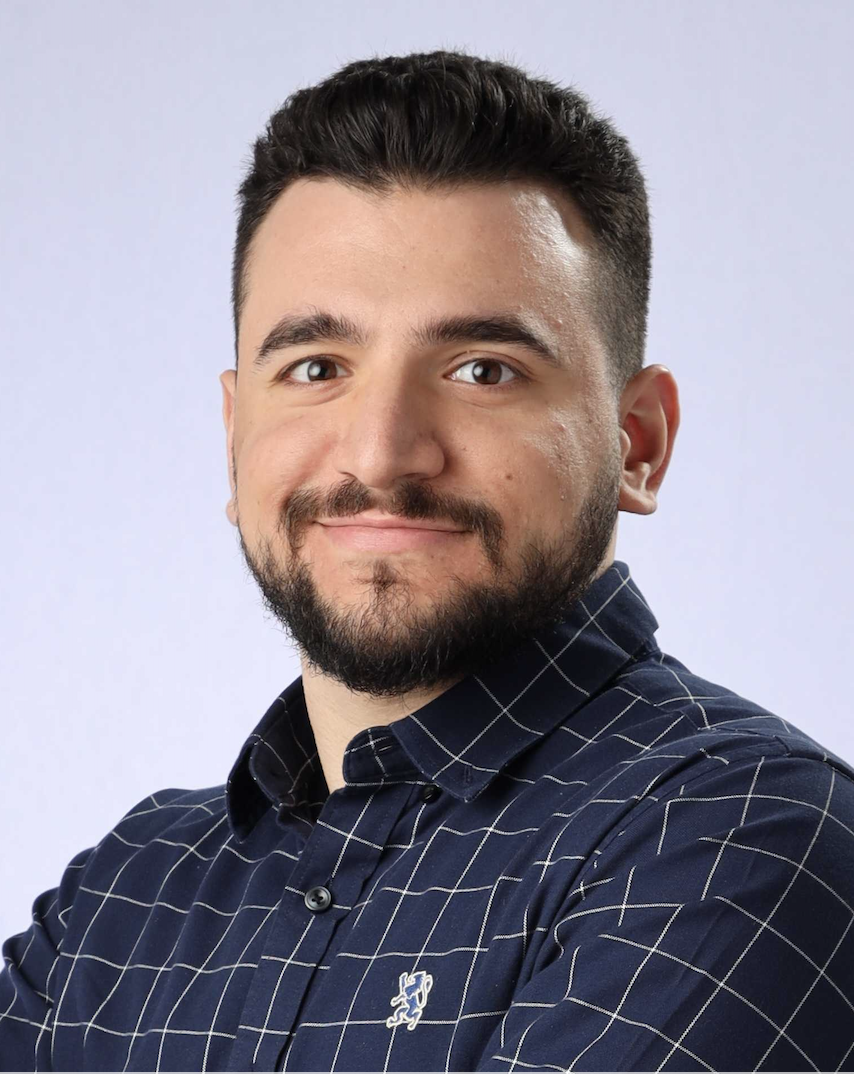}}]
{Mohammad Dib}
\textit{(mohammad.dib@multiverse
computing.com)}
received his BSc in Electrical Engineering from the American University of Sharjah and his MASc in Machine Learning from the University of Waterloo. He joined Multiverse Computing in 2022 as a Machine Learning Engineer, contributing to projects across multiple domains, including finance and manufacturing.
\end{IEEEbiography}

\begin{IEEEbiography}[{\includegraphics[width=\biowidth,trim=0.2in 0in 0.2in 0in,clip]{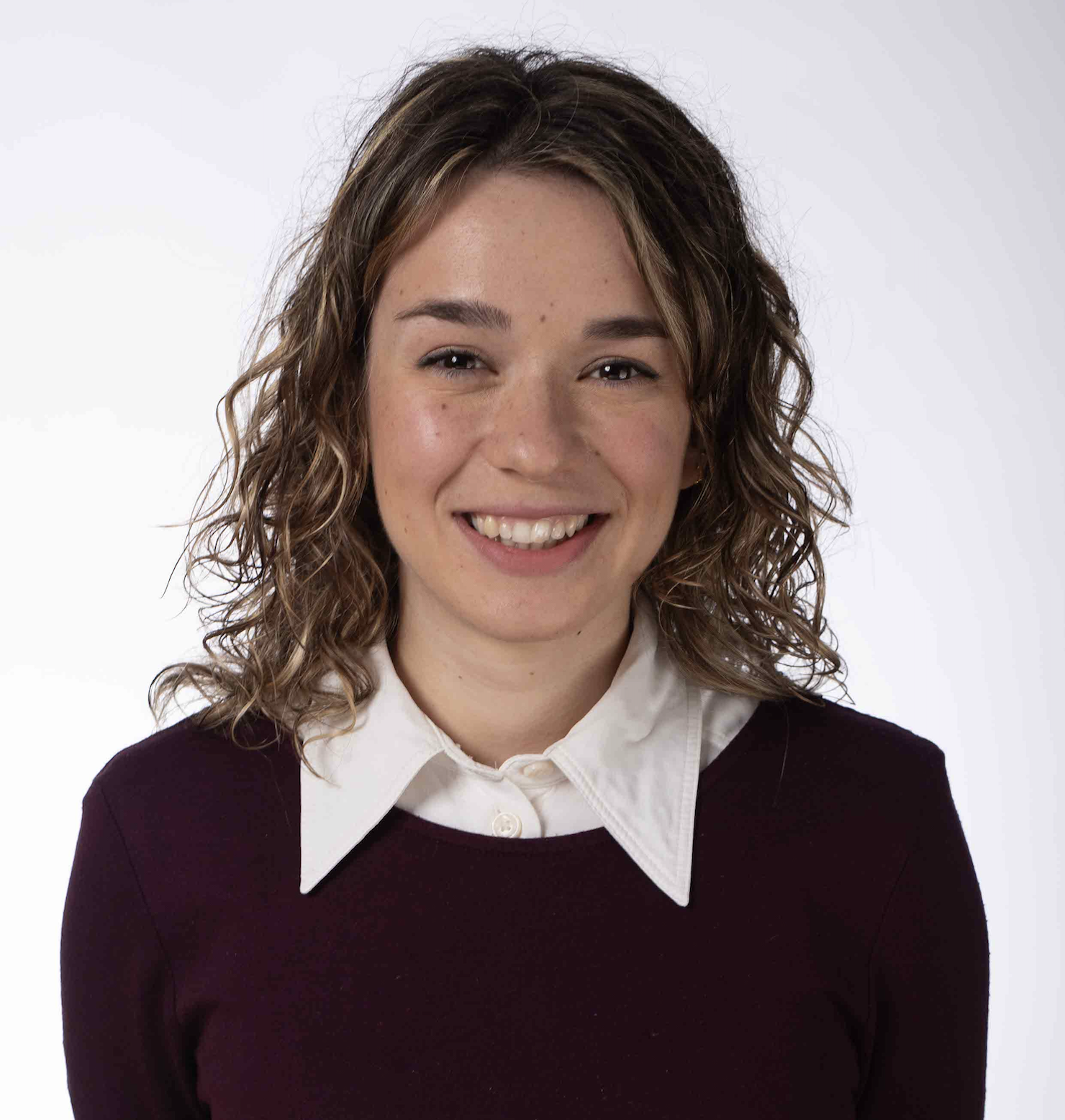}}]
{Alba Villarino}
\textit{ (alba.villarino@multiversecompu ting.com)}
received her degree in Physics from the University of Salamanca and her Master’s degree in Quantum Science and Technologies from the University of the Basque Country (UPV/EHU) in 2021. She has over three years of experience in the field of quantum computing. She previously worked as a Quantum Software Engineer and is currently a Manager at Multiverse Computing.
\end{IEEEbiography}

\begin{IEEEbiography}[{\includegraphics[width=\biowidth,trim=0.9in 0in 1.4in 0in,clip]{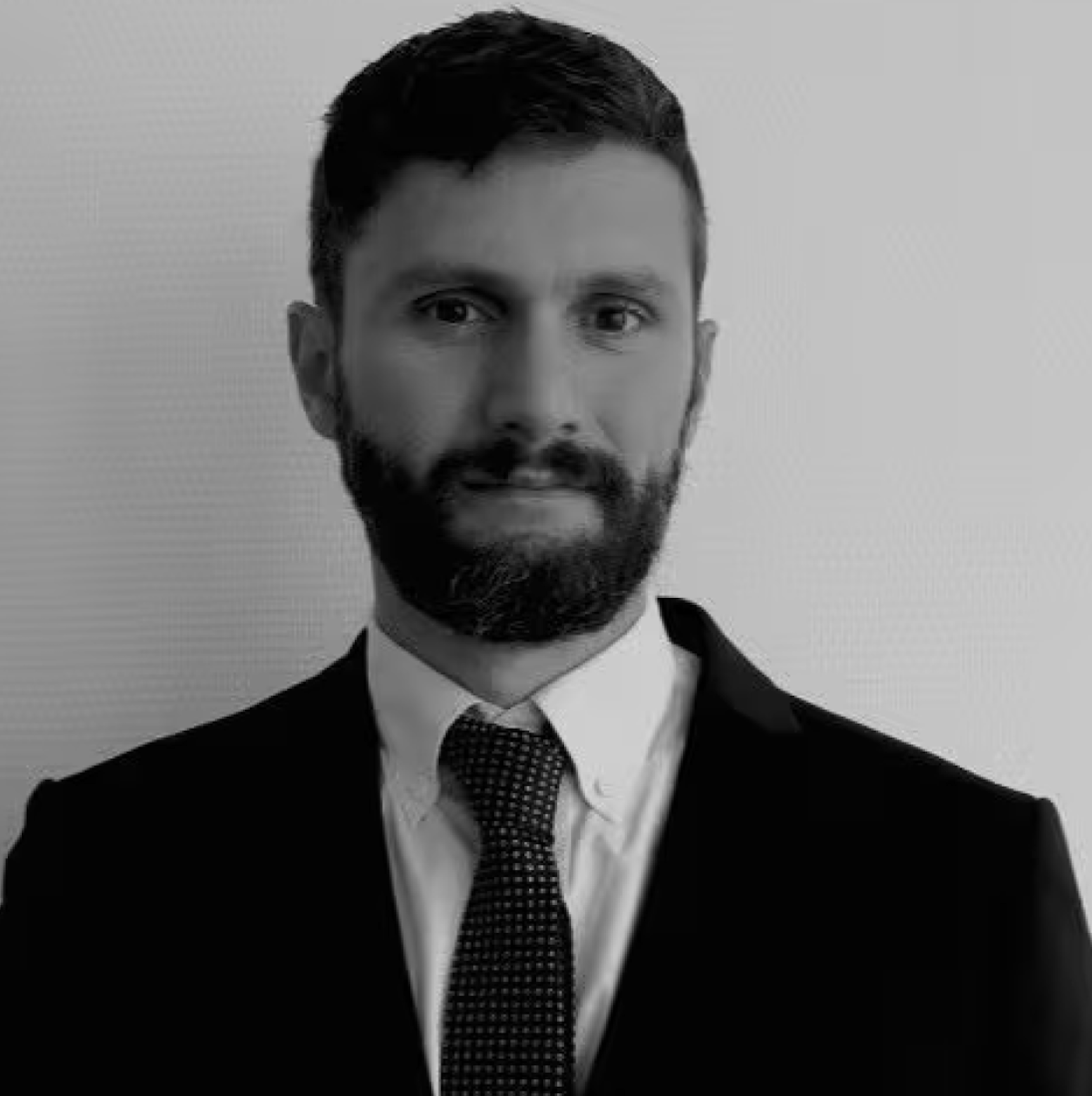}}]
{Aser Cortines}
\textit{ (aser.cortines@multiversecompu ting.com)}
holds an M.Sc. from École Polytechnique and a Ph.D. in Applied Mathematics from the University of Paris Diderot. He held postdoctoral roles at Technion and the University of Zurich. His experience includes quantitative modeling in energy markets, banking model risk assessment, and financial engineering. Since joining Multiverse Computing in 2022, he has led projects applying quantum technologies, and currently serves as Director of Engineering, overseeing work in finance, optimization, and machine learning.
\end{IEEEbiography}

\begin{IEEEbiography}[{\includegraphics[width=\biowidth,trim={1.25in 0in 1.25in 0in},clip]{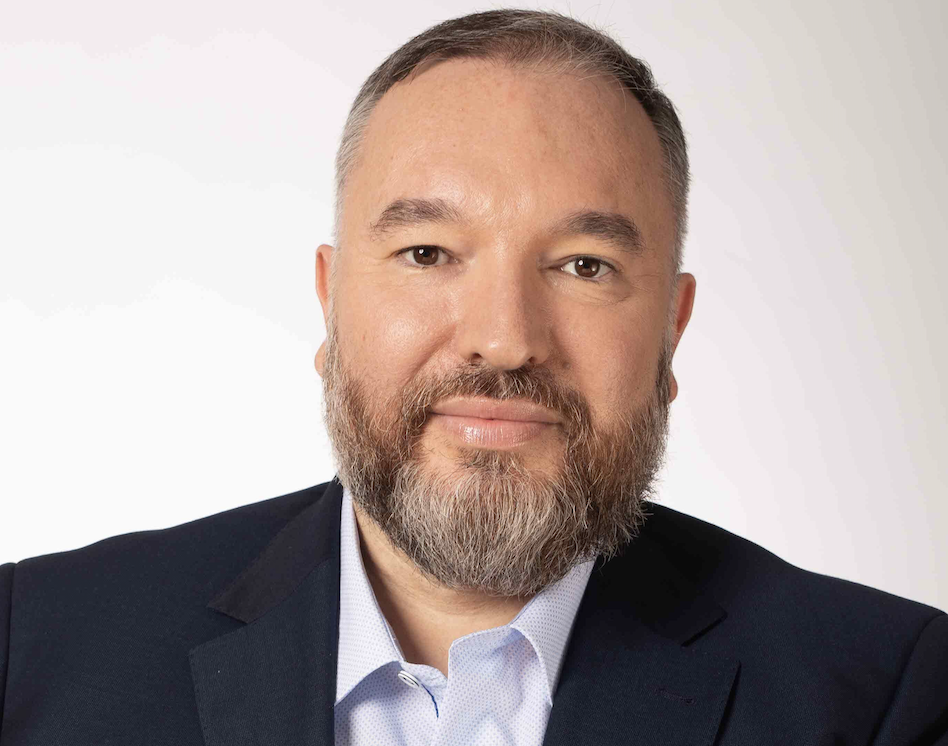}}]
{Román Or\'us}
\textit{ (roman.orus@multiversecomputing .com)}
Ph.D. in quantum physics from the University of Barcelona. Ikerbasque Research Professor at the Donostia International Physics Center (DIPC). President of the Specialized Group on Quantum Information (GEIC) and of the Spanish Royal Physical Society (RSEF). Early Career Prize from the European Physical Society, and Marie Curie Fellow. Editor of the journal QUANTUM. Partner at Entanglement Partners. Roman is in charge of the scientific direction of the company (CSO). Research and development of new quantum algorithms, as well as new TN algorithms, both for optimization and machine learning. He also coordinates the scientific team, identifies and hires key technical staff, and is the main responsible for writing scientific papers and reports. He develops networking with hardware vendors and other quantum software companies.
\end{IEEEbiography}

\end{document}